\documentclass[runningheads]{llncs}
\usepackage[T1]{fontenc}
\usepackage{graphicx}
\usepackage[hidelinks]{hyperref}
\usepackage{chronology}
\usepackage{amssymb}
\usepackage{physics}
\usepackage{braket}
\usepackage{mathtools}

\begin{document}
\title{On the Theory of Quantum and Towards Practical Computation
%\thanks{Supported by organization x.}
}

\author{Robert Kudelić
\inst{}
%\orcidID{0000-1111-2222-3333} 
%	\and Second Author\inst{2,3}\orcidID{1111-2222-3333-4444} \and Third Author\inst{3}\orcidID{2222--3333-4444-5555}
}

\authorrunning{Robert Kudelić et al.}
\institute{University of Zagreb Faculty of Organization and Informatics,\\ Republic of Croatia
%\and Springer Heidelberg, Tiergartenstr. 17, 69121 Heidelberg, Germany
}
\maketitle              % typeset the header of the contribution
\begin{abstract}
Quantum computing exposes the brilliance of quantum mechanics through computer science and, as such, gives oneself a marvelous and exhilarating journey to go through. This article leads along that journey with a historical and current outlook on quantum computation that is geared toward computer experts but also to experts from other disciplines as well. It is an article that will bridge the vast gap between classical and quantum computation and open an entering wedge through which one will be able to both bring himself up to speed on quantum computation and, intrinsically, in a straightforward manner, become acquainted with it. We are indeed in luck to be living in an age where computing is being reinvented, and not only seeing history in the making firsthand but, in fact, having the opportunity to be the ones who are reinventing--and that is quite a thought.

\keywords{Quantum Computation \and Fundamentals \and Review \and History \and Open Questions \and Quantum Phenomena \and Technology \and Algorithm Design Pattern \and Application.}
\end{abstract}
\section{For Once It All Began}
%    Towards Quantum Computation
\label{sec:History}
How vast the chasm is, how difficult it is to grasp it, and how steep the learning curve has become--and perhaps always has been--is a realization to which one arrives when, for the first time, tries to bring oneself to a destination called quantum (QTM) computation. It is an awe-inspiring journey that through this article we will relive, unsealing its complex secrets, and gradually grasping computation known as quantum computation\footnote{With this paper, we will try to complete the picture on quantum computation for interested parties that are laying outside of physics and at the same time give the reader both a review and the state of the art in a manner different from that of classical review papers.}.

\begin{figure}
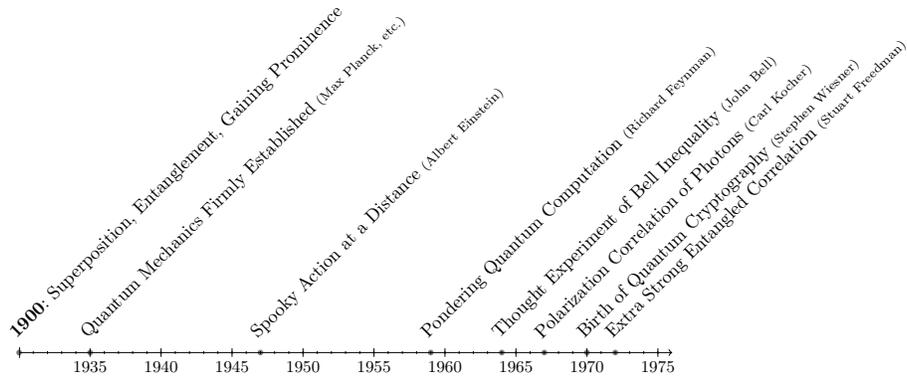

	\begin{chronology}[5]{1930}{1975}{\textwidth}
		
		% stavljeno 1930 radi preglednosti, naznačeno u tekstu kako je 1900
		\event{1930}{{\normalsize \textbf{1900}: Superposition, Entanglement, Gaining Prominence}}
		
		\event{1935}{{\normalsize Quantum Mechanics Firmly Established} {\scriptsize (Max Planck, etc.)}}
		
		\event{1947}{{\normalsize Spooky Action at a Distance} {\scriptsize (Albert Einstein)}}
		
		\event{1959}{{\normalsize Pondering Quantum Computation} {\scriptsize (Richard Feynman)}}
		
		\event{1964}{{\normalsize Thought Experiment of Bell Inequality} {\scriptsize (John Bell)}}
		
		\event{1967}{{\normalsize Polarization Correlation of Photons} {\scriptsize (Carl Kocher)}}
		
		\event{1970}{{\normalsize Birth of Quantum Cryptography} {\scriptsize (Stephen Wiesner)}}
		
		\event{1972}{{\normalsize Extra Strong Entangled Correlation} {\scriptsize (Stuart Freedman)}}
		
	\end{chronology}
	\caption{Quantum Computing Timeline: 1900-1979.} \label{kron:1900}
\end{figure}

\begin{figure}
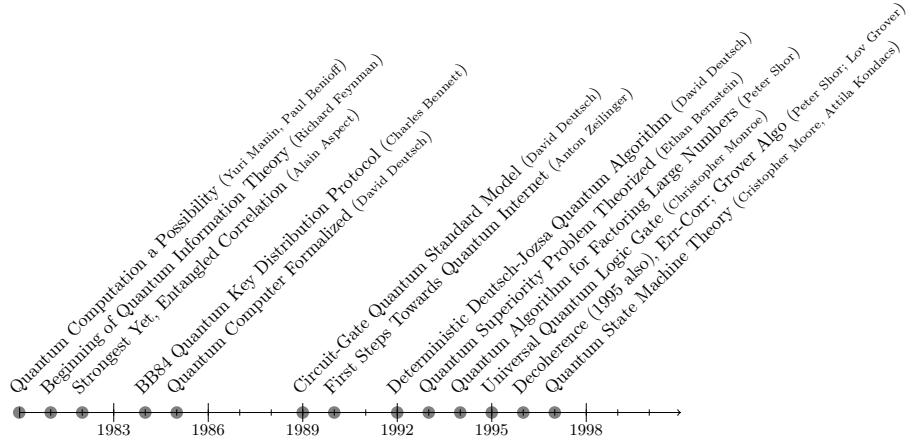

	\begin{chronology}[3]{1980}{2000}{\textwidth}
		
		\event{1980}{{\normalsize Quantum Computation a Possibility} ({\scriptsize Yuri Manin, Paul Benioff})}
		
		\event{1981}{{\normalsize Beginning of Quantum Information Theory} ({\scriptsize Richard Feynman})}
		
		\event{1982}{{\normalsize Strongest Yet, Entangled Correlation} ({\scriptsize Alain Aspect})}
		
		\event{1984}{{\normalsize BB84 Quantum Key Distribution Protocol} ({\scriptsize Charles Bennett})}
		
		\event{1985}{{\normalsize Quantum Computer Formalized} ({\scriptsize David Deutsch})}
		
		\event{1989}{{\normalsize Circuit-Gate Quantum Standard Model} ({\scriptsize David Deutsch})}
		
		\event{1990}{{\normalsize First Steps Towards Quantum Internet} ({\scriptsize Anton Zeilinger})}
		
		\event{1992}{{\normalsize Deterministic Deutsch-Jozsa Quantum Algorithm} ({\scriptsize David Deutsch})}
		
		\event{1993}{{\normalsize Quantum Superiority Problem Theorized} ({\scriptsize Ethan Bernstein})}
		
		\event{1994}{{\normalsize Quantum Algorithm for Factoring Large Numbers} ({\scriptsize Peter Shor})}
		
		\event{1995}{{\normalsize Universal Quantum Logic Gate} ({\scriptsize Christopher Monroe})}
		
		\event{1996}{{\normalsize Decoherence (1995 also), Err-Corr; Grover Algo} ({\scriptsize Peter Shor; Lov Grover})}
		
		\event{1997}{{\normalsize Quantum State Machine Theory} ({\scriptsize Cristopher Moore, Attila Kondacs})}
		
		%        \event{}{{\normalsize } ({\scriptsize })}
		%\event[1915]{1918}{two}
		%\event{\decimaldate{25}{12}{1925}}{three}
	\end{chronology}
	\caption{Quantum Computing Timeline: 1980-2000.} \label{kron:1980}
\end{figure}

\begin{figure}
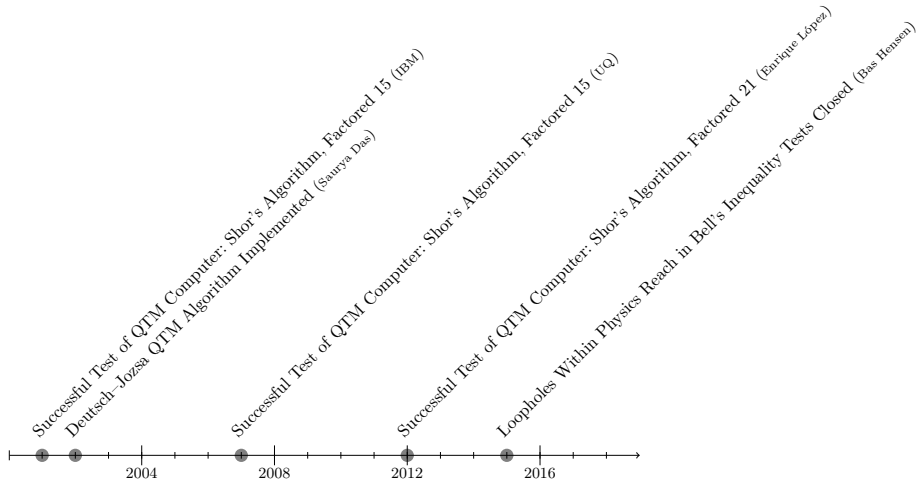

	\begin{chronology}[4]{2000}{2018}{\textwidth}
		
		\event{2001}{{\normalsize Successful Test of QTM Computer: Shor's Algorithm, Factored $ 15 $} ({\scriptsize IBM})}
		
		\event{2002}{{\normalsize Deutsch–Jozsa QTM Algorithm Implemented} ({\scriptsize Saurya Das})}
		
		\event{2007}{{\normalsize Successful Test of QTM Computer: Shor's Algorithm, Factored $ 15 $} ({\scriptsize UQ})}
		
		\event{2012}{{\normalsize Successful Test of QTM Computer: Shor's Algorithm, Factored $ 21 $} ({\scriptsize Enrique López})}
		
		\event{2015}{{\normalsize Loopholes Within Physics Reach in Bell's Inequality Tests Closed} ({\scriptsize Bas Hensen})}

		%\event[1915]{1918}{two}
		%\event{\decimaldate{25}{12}{1925}}{three}
	\end{chronology}
	\caption{Quantum Computing Timeline: 2001-2015.} \label{kron:2001}
\end{figure}

\begin{figure}
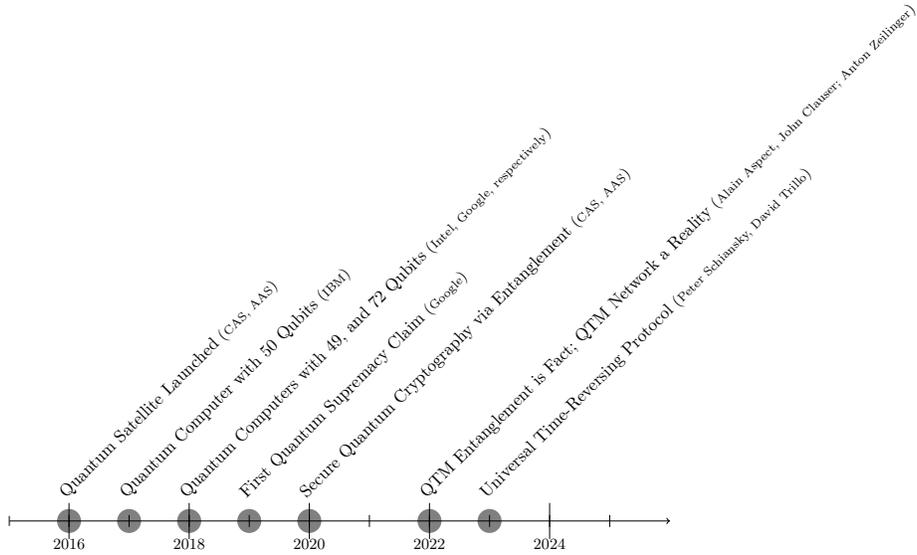

	\begin{chronology}[2]{2015}{2025}{\textwidth}
		
		\event{2016}{{\normalsize Quantum Satellite Launched} ({\scriptsize CAS, AAS})}
		
		\event{2017}{{\normalsize Quantum Computer with $ 50 $ Qubits} ({\scriptsize IBM})}
		
		\event{2018}{{\normalsize Quantum Computers with $ 49 $, and $ 72 $ Qubits} ({\scriptsize Intel, Google, respectively})}
		
		\event{2019}{{\normalsize First Quantum Supremacy Claim} ({\scriptsize Google})}
		
		\event{2020}{{\normalsize Secure Quantum Cryptography via Entanglement} ({\scriptsize CAS, AAS})}
		
		\event{2022}{{\normalsize QTM Entanglement is Fact; QTM Network a Reality} ({\scriptsize Alain Aspect, John Clauser; Anton Zeilinger})}
		
		\event{2023}{{\normalsize Universal Time-Reversing Protocol} ({\scriptsize Peter Schiansky, David Trillo})}

		%\event[1915]{1918}{two}
		%\event{\decimaldate{25}{12}{1925}}{three}
	\end{chronology}
	\caption{Quantum Computing Timeline: 2016-2023.} \label{kron:2016}
\end{figure}

Before we therefore begin with the subject at hand, it would be of interest to give a brief historical background and a more forward motivation behind this work\footnote{In Figures \ref{kron:1900}, \ref{kron:1980}, \ref{kron:2001}, and \ref{kron:2016}, one can visually grasp the quantum timeline as it relates to quantum computing, with a number of milestones presented.}. It all began long ago, perhaps some years before what is typically remembered. All the way back in 1935, the principles of quantum mechanics\footnote {Quantum communication at the theoretical level was proposed by Albert Einstein. \cite{Chen2021}} where already heavily discussed \cite{Bohr1935,Einstein2004}, namely superposition (particle being in multiple states at the same time, until observed \cite{Dowling2003}) and entanglement (correlation between particle states no matter the distance between them \cite{Dowling2003}), which we will soon define in more detail, that are so crucial to quantum computation as well \cite{AHARONOV1999}. A number of decades prior to those events, on December 14, 1900, to be exact, Max Planck struck the beginning of quantum mechanics "at a meeting of the German Physical Society". \cite{Passon2017} Those were tumultuous and exciting days, I presume\footnote{Prior to Planck's reveal of his today known Planck’s law, the existence of the atom was scientifically debated and established \cite{Grossman2017,Ball2016}, after which Ludwig Boltzmann in 1872 suggested that small particles could have multiple energy levels instead of the one being observed \cite{Flamm1983}--with Boltzmann substantially influencing both science and later works of scientists such as Max Planck and Albert Einstein. \cite{Flamm1983}}, but the best was yet to come. A few decades have passed, and ideas and research were advancing to and fro. Some scientists, excited, trying to advance the theory of quantum mechanics, while others were working against it, but not only against it, even fighting it\footnote{It is famously remembered in science how Einstein, who himself had doubts about quantum phenomena, told Max Born in 1947 that quantum mechanics entanglement of particles represents "spooky action at a distance." \cite{Einstein2004,Boughn2022}}--which in science is business as usual: That which nature's physical systems deny, needs to perish.

Then one day, as the knowledge increased, some started pondering about computation that is microscopic and able to simulate physical systems with which classical computers have difficulty. \cite{Preskill2021} That person, right at the forefront, thinking these "microscopic" thoughts that were far beyond the abilities of those days was Richard Feynman. \cite{Preskill2021,Albash2018} It is not known when exactly he first started pondering the idea of a quantum computer, but what is known is that in his 1959 talk, he was predicting an enormous miniaturization of technology, even to the size of an atom. \cite{Preskill2021,Albash2018} There was nothing that he saw in the laws of nature that wouldn't allow this miniaturization, and he was speaking about it. \cite{Preskill2021,Albash2018} Time has passed, and Feynman, together with other scientists, tried to advance the issue. Then something happened, and a theory so necessary for practical quantum computation started to emerge\footnote{In 1973, C. H. Bennett established logical reversibility of computation, where any Turing machine, "general-purpose computing automaton", can be "made logically reversible at every step" \cite{Bennett1973}--this find is also important for quantum computation in terms of its own logical and physical reversibility. \cite{Qiu2008,Ciamarra2001}}.

In 1981, Feynman gave a conference talk\footnote{Feynman was pointing out that classical computers, which were then still in their infancy, are simply inadequate to succinctly describe the "quantum state of many particles" \cite{Preskill2021}. If one thinks about it, at a beginning of the digital age we live in, he was already calling for the next revolution in computing and projecting it onto its natural application, simulation of the world we live in at the most fundamental level. \cite{Preskill2021}} on "simulating physics with computers"\footnote{Which are his own words recorded in a journal publication of the same title, in 1982. \cite{Feynman1982}} \cite{Preskill2021,Chen2021}, which was later published as an edited transcript \cite{Preskill2021} in a scientific journal \cite{Feynman1982}--and for all intents and purposes this event launched "quantum computing as a field of study" \cite{Preskill2021,Albash2018}, "which established the beginning of quantum information theory" \cite{Chen2021}. At about the same time, others were investigating as well, and from then onward, nothing was ever the same. What is fascinating is that both Manin \cite{Manin1980} and Benioff \cite{Benioff1980} were just a year prior, in 1980, bringing into the foreground ideas of large significance. Manin was in his book Computable and Uncomputable \cite{Hidary2021,Preskill2021} discussing how simulating a many-particle system requires exponential cost on a classical computer \cite{Preskill2021,Manin1980,Albash2018}, while Benioff went further down the quantum line, complementing Manin, in explaining how one would describe computation from the quantum outlook and suggesting by the construction of such a model that quantum computation might be a possibility \cite{Benioff1980}.

On a bit different note, the question that was continually puzzling Einstein, whether two particles really can be entangled and have correlation between their states without a hidden information, was being experimentally answered by Alain Aspect et al., and the answer was yes\footnote{In a 1982 paper titled "Experimental Realization of Einstein-Podolsky-Rosen-Bohm Gedankenexperiment: A New Violation of Bell's Inequalities" \cite{Aspect1982}, it was shown that experiments do not support hidden variable theory promoted by Einstein and others; Bell's inequality is not satisfied, and therefore there is no hidden information that would explain quantum entanglement \cite{Aspect1982,Chen2021}--Bell's inequality is a thought experiment published in 1964 by John Stewart Bell in order to test Einstein's idea. The inequality states that if hidden variables are real, then correlation between the properties of particles is happening, but only to a certain degree. \cite{Bell1964}}, they can. \cite{Chen2021} With the first real-world experiments that were successful and conducted a decade earlier, in 1972 by Freedman and Clauser (which in turn depended on the work of Carl Kocher \cite{Kocher1967}), and with an extra-strong correlation being observed \cite{Cho2022,Freedman1972}, Aspect's work, where "the greatest violation of generalized Bell's inequalities" \cite{Aspect1982} ever was achieved, has put the predictions of quantum mechanics strongly on the map.

Only a few years later, in 1985, another important advance came when David Deutsch\footnote{Known for the deterministic Deutsch-Jozsa quantum algorithm published in 1992 \cite{Deutsch1992}, with improvement and implementation following in 1998 \cite{Collins1998} and 2002. \cite{Das2002,Wei2006}} "formalized the notion of a quantum computer" \cite{Preskill2021,Deutsch1985} and raised the question: "Whether quantum computers might have an advantage over classical computers at solving problems that have nothing to do with quantum physics" \cite{Preskill2021,Deutsch1985}. True, the algorithm that Deutsch and Jozsa later published \cite{Deutsch1992} was of little practical significance, but it showed superiority in efficiency of the quantum algorithm over its deterministic classical counterpart. \cite{Deutsch1992,Simon1994} Thinking about quantum computation and ideas that came through Benioff \cite{Benioff1980} and Feynman \cite{Feynman1982} Deutsch was led to in 1989\footnote{Also the year when quantum key distribution protocol was implemented for the first time, while the distance on which it was transmitted was less than one meter--"transmission range is mainly bounded by the damping of light signals in fiber-optical cables, loss of photons, and also external noises." \cite{Savchuk2019}} propose what later became the standard model for describing quantum computation, the well-known circuit-gate model \cite{Deutsch1989}. \cite{Albash2018}

With Deutsch formalizing the notion of quantum computer, Umesh Vazirani and his student Ethan Bernstein were formulating "a contrived\footnote{With the research showing that by the artificial problem devised one violates complexity-theoretic Church-Turing thesis \cite{Bernstein1997} which states that any computation model can be simulated by probabilistic Turing machine in polynomial time \cite{Robic2020}--the proof for this, however, is out of sight and difficult to obtain, unless some revolutionary breakthrough in complexity theory occurs. \cite{Bernstein1997,Vazirani2002}} problem that a quantum computer could solve with a super-polynomial speedup over a classical computer" \cite{Preskill2021,Bernstein1997}--that was in 1993\footnote{In the same year, another important result was obtained; it was demonstrated that "any function computable in polynomial time by a quantum Turing machine has a polynomial-size quantum circuit" \cite{Yao1993}--this result enabled the construction of a universal quantum computer "which can simulate, with a polynomial factor slowdown, a broader class of quantum machines than that of" Bernstein et al. \cite{Yao1993,Bernstein1993}} \cite{Bernstein1993}, that is\footnote{Approximately a decade later from the work of Aspect \cite{Aspect1982}, at the beginning of the 1990s, Zeilinger's group was working on swapping and extending entanglement to distant particles \cite{Cho2022,Greenberger1990,Zukowski1993}, these steps were the first taken towards quantum internet \cite{Cho2022}--Alain Aspect, John F. Clauser, and Anton Zeilinger, together with men that worked with them on entanglement and communicating quantum information, have pioneered Quantum Information Science. \cite{NobelPhysics2022}}. The same superiority was presented in 1994 by Daniel Simon, who showed that by solving the idealized version of the problem, which is finding the function period, quantum computers could indeed achieve an exponential improvement in speed when compared to their classical counterparts. \cite{Simon1994,Simon1997,Preskill2021} And despite the fact that Simon's idea, just like the one from Deutsch, had little practical weight and no application in sight, that was soon to change, for in just a short while, tremendous happenings will occur for quantum computation. \cite{Preskill2021}

The same idea and an instance where quantum computers would show their superiority has, in 1994, inspired Peter Shor to baffle the world and publish the paper in which he presented an efficient way for Fourier transform calculation, which he used for a definition of an efficient algorithm for computing discrete logarithms--and all this was done for a quantum computer. \cite{Shor1994,Shor1999,Preskill2021} But that was not the end. A few days later, after the aforementioned breakthrough, and by using similar ideas \cite{Preskill2021}, in the same seminal paper, Shor presented "an efficient quantum algorithm for factoring large numbers" \cite{Shor1994,Preskill2021}. \cite{Shor1994,Shor1999} The implications for cryptanalysis were enormous\footnote{It is a well-known fact that today's asymmetric key computer cryptography is based on large semi-prime number factorization \cite{Rivest1983,Rivest1977}, and Shor's quantum algorithm for prime factorization therefore created quite a commotion \cite{Preskill2021,Shor1999}--as a fascinating digression, the well-known public key cryptography was not actually first invented in 1977 by Rivest, Shamir, and Adleman \cite{Rivest1977}, but by Clifford Cocks (an employee of the British intelligence agency) in 1973, based on the work of his colleges at work (Ellis and Williamson), a story kept secret for 24 years and revealed in 1977 at a conference, supported by Government Communications Headquarters (the British signals intelligence agency) internal declassified documents. \cite{Smart2008}}, and the interest in quantum computing has once again exploded. \cite{Preskill2021}

All was not well in the land called Q-Country, though, and at the same time those great achievements were being made, a dark cloud was looming over quantum computation, and that dark cloud was called decoherence\footnote{More on decoherence in the section on foundational terminology.}--an inability for a computer to compute in a quantum manner because of interaction with the outside world\footnote{It is remembered to be said of quantum computation in those days: "In this sense the large-scale quantum machine, though it may be the computer scientist's dream, is the experimenter's nightmare." \cite{Haroche1996}}. \cite{Preskill2021,Landauer1995,Unruh1995,Haroche1996} But the question of decoherence was already being tackled and is one of the main issues with quantum hardware that remains to be tackled to this day. \cite{McEwen2021,CampagneIbarcq2020,Choi2020,Cai2021} Shor himself has already, in 1995 and 1996, published research on quantum error-correcting codes and on fault-tolerant methods by which one could compute on quantum hardware, which is rather noisy, in a reliable manner. \cite{Shor1995,Steane1996,Shor1996,Preskill2021} And with that, "by the end of 1996 it was understood, at least in principle, that quantum computing could be scaled up to large devices that solve very hard problems, assuming that errors afflicting the hardware are not too common or strongly correlated" \cite{Preskill2021,Aharonov1997,Knill1998,Preskill1998}--which is confirmed by the latest research dealing with quantum computation, scalability, and decoherence\footnote{The “accuracy threshold theorem” for quantum computing has very rapidly seen daylight, only $2\frac{1}{2}$ years after Shor discovered his algorithm. \cite{Preskill2021}}: "fault-tolerant quantum computation will be practically realizable." \cite{Krinner2022}.

During those same exciting times \cite{Preskill2021}, as John Preskill adequately called them \cite{Preskill2021,Preskill1998a}, another important realization was happening. It was the year 1995 when Cirac and Zoller published that, with the tools in atomic physics and quantum optics, one could implement a quantum computer and perform quantum logical operations. \cite{Cirac1995} Building on that foundation, a few months later in the same year, Monroe et. al. demonstrated a fundamental quantum logic gate, "operation of a two-bit controlled-NOT quantum logic gate", to be exact \cite{Monroe1995}, which, coupled with simple single-bit operations, formed a universal quantum logic gate\footnote{In 2012, David Wineland and Serge Haroche won a Nobel Prize in Physics for their work on microscopic objects and the effects of their manipulation \cite{Schirber2012,NPrize2012}--this was one out of numerous Nobel prizes given for accomplishments that are linked to quantum effects (the following compact list generally excludes fluids): 1918 (Max Karl Ernst Ludwig Planck) \cite{NPrize1918}, 1919 (Johannes Stark) \cite{NPrize1919}, 1921 (Albert Einstein) \cite{NPrize1921}, 1923 (Robert Andrews Millikan) \cite{NPrize1923}, 1927 (Arthur Holly Compton) \cite{NPrize1927}, 1929 (Louis-Victor Pierre Raymond de Broglie) \cite{NPrize1929}, 1932 (Werner Karl Heisenberg) \cite{NPrize1932}, 1933 (Erwin Schrödinger, Paul A. M. Dirac) \cite{NPrize1933}, 1937 (Clinton Joseph Davisson, George Paget Thomson) \cite{NPrize1937}, 1954 (Max Born, Walther Bothe) \cite{NPrize1954}, 1964 (Charles Hard Townes, Nicolay Gennadiyevich Basov, Aleksandr Mikhailovich Prokhorov) \cite{NPrize1964}, 1965 (Sin-Itiro Tomonaga, Julian Schwinger, Richard P. Feynman) \cite{NPrize1965}, 1972 (John Bardeen, Leon Neil Cooper, John Robert Schrieffer) \cite{NPrize1972}, 1973 (Leo Esaki, Ivar Giaever, Brian David Josephson) \cite{NPrize1973}, 1978 (Pyotr Leonidovich Kapitsa) \cite{NPrize1978}, 1979 (Sheldon Lee Glashow, Abdus Salam, Steven Weinberg) \cite{NPrize1979}, 1981 (Nicolaas Bloembergen, Arthur Leonard Schawlow, Kai M. Siegbahn) \cite{NPrize1981}, 1984 (Carlo Rubbia, Simon van der Meer) \cite{NPrize1984}, 1985 (Klaus von Klitzing) \cite{NPrize1985}, 1987 (J. Georg Bednorz, K. Alexander Müller) \cite{NPrize1987}, 1989 (Hans G. Dehmelt, Wolfgang Paul) \cite{NPrize1989}, 1998 (Robert B. Laughlin, Horst L. Störmer, Daniel C. Tsui) \cite{NPrize1998}, 1999 (Gerardus‘t Hooft, Martinus J. G. Veltman) \cite{NPrize1999}, 2005 (Roy J. Glauber, John L. Hall, Theodor W. Hänsch) \cite{NPrize2005}, 2012 (Serge Haroche, David J. Wineland) \cite{NPrize2012}, 2022 (Alain Aspect, John F. Clauser, Anton Zeilinger) \cite{NPrize2022}.} \cite{Monroe1995}--this was quite important piece of the quantum computing puzzle, since if correct and practical model of computation can not be found, then all efforts, perspiration and tears would be in vain. With previous breakthroughs, especially those that happened during the last decennia of the 20th century, a strong foundation was laid, and it seemed quite possible that one day quantum computation would be a reality. The possibility of that reality was never given up, and research continued.

Then, soon after Shor presented his Las Vegas quantum algorithms \cite{Shor1994}, in 1996 Lov Grover emerged with another fascinating discovery: it is possible to search a database for an entry in $ \sqrt{N} $ time and identify a record with a probability of $ \frac{1}{2} $ \cite{Grover1996}--which then represents quantum Monte Carlo, and is asymptotically optimal \cite{Grover1998}, and by repeated sampling, this probability can arbitrarily grow \cite{Grover1996}. A substantial achievement since classical machines, both deterministic and probabilistic, will need $ \frac{N}{2} $ time to achieve the same probability bound of $ \frac{1}{2} $, and only in an ordered list via Binary search can classical machines achieve $\log_2 N $ time. \cite{Grover1997a} Some, as well researching in quantum computing, were advancing tools for better understanding such computations and developing theories for quantum state machines, 1997 was the year. \cite{Moore1997,Moore2000,Kondacs1997}

Shortly after, just a few years have passed, in 2001, the company was IBM, and scientists there have announced that successful testing of a quantum computer has been conducted. The capacity of the machine was $ 7 $ qubits (first register $ 3 $, second register $ 4 $), and the quantum computer itself was implemented by nuclear magnetic resonance \footnote{Nuclear magnetic resonance is defined by "selective absorption of very high-frequency radio waves by certain atomic nuclei that are subjected to an appropriately strong stationary magnetic field" \cite{EncyclopaediaBritannica2023}--for details, one can look in \cite{Hoult1997}.}. \cite{Savchuk2019} Shor's algorithm was executed on this machine, and by employing quantum effects, number $ 15 $ was factorized \cite{Savchuk2019}--this achievement was for the history books, deserving of noting big success. Then again, in 2007\footnote{Similar experiment was carried out at the University of Science and Technologies of China, this time $ 6 $ qubits were used (first register $ 2 $, second register $ 4 $). \cite{Savchuk2019}}, a validation came when scientists at the University of Queensland (UQ) experimentally demonstrated execution of Shor's algorithm for large number factorization by "using quantum logic gates based on photon polarization"--they have also factorized number $ 15 $ (first register $ 3 $ qubits, second register $ 4 $ qubits). \cite{Savchuk2019} At this stage, quantum computation has gone from theory to practice. By the end of the 1990s, enough foundational theory had been discovered, and the beginning of the 21st century was the dawn of practical quantum computation. Machines are being built, and algorithms are being implemented\footnote{For a short insight into quantum computers of those days, one can consult \cite{Blatt2008,Thompson2008}.}, and now theory and practice go together.

And so in 2009 and 2012\footnote{Company D-Wave Systems claimed in 2012 a construction of a quantum device with $ 84 $ qubits, then in the same year, a $ 512 $ qubit quantum computer was announced, while in 2015 a creation of a $ 1152 $ qubit quantum computer was stated. \cite{Savchuk2019} There is a debate, though, about whether these computers are quantum or not, since, for example, algorithms like Simon and Bernstein-Vazirani can be run on them while others like Grover and Shor cannot. \cite{Savchuk2019} Researchers at Google in 2015 claimed that these devices do use quantum effects \cite{Savchuk2019}, but is that enough for a device to be called a quantum computer? After analysis of available information on D-Wave devices, it was concluded that they "do not provide any computational advantage over the classical computer", calling it a quantum annealer \cite{Savchuk2019}--it is possible that experiments were testing world-class skiers on a bunny slope course; time will tell. \cite{Cho2014}} new experiments have confirmed the reality of quantum computation, making it even stronger; one more successful experimental demonstration of Shor's algorithm has taken place, the method was an integrated wave-guide based on a silicon chip, with only $ 4 $ qubits based on photons used for factorization of number $ 15 $ (first register $ 1 $ qubit, second register $ 3 $ qubits). \cite{Savchuk2019} And as a supplementation, in 2012, at the University of California (UC), one more experiment successfully factored number $ 15 $, Shor's algorithm in action, "using phase qubits and superconducting wave resonators", with $ 4 $ qubits, just like the previous group of researchers (but in the first register there were $ 2 $ qubits, and in the second $ 2 $ as well). \cite{Savchuk2019}

This series of implementations of quantum computers and successful algorithm runs continued, and soon there was quite a group of scientists that have dabbled in quantum computing and have witnessed its strangeness and marvelousness at the same time, e.g. Martin-Lopez et al. in \cite{MartinLopez2012} with factoring number $ 21 $, via Shor, "using only two photon-based qubits" (2012), Nanyang Xu et al. in \cite{Xu2012} turning factorization problem into optimization problem, by a scheme\footnote{Improved in \cite{Schaller2010} by Gernot Schaller and Ralf Schutzhold.} from Burges from Microsoft Research, and factoring number $ 143 $\footnote{In \cite{Dattani2014} it has been demonstrated that larger numbers have been factored without authors knowing, e.g. $ 56153 $, and in order to exploit the power of quantum computers, the authors have discussed scheme with more qubits to solve discrete optimization problem, an example of factoring $ 291311 $ with $ 6 $ qubits was given. \cite{Dattani2014} The paper has also made a demonstration of quantum factorization of triprime $ 175 $ with $ 3 $ qubits, a task difficult for classical factorization algorithms but relatively easy for a quantum algorithm. \cite{Dattani2014}} with $ 4 $ qubits only, this was an adiabatic algorithm run on a liquid crystal nuclear magnetic resonance quantum processor, and for example, Thomas Monz et al. in \cite{Monz2016}, via five trapped calcium ions on a quantum computer, implemented a scalable version of Shor's algorithm, with the approach providing "potential for designing a powerful quantum computer, but with fewer resources." \cite{Savchuk2019}

Next, it was to IBM again, which has seriously grabbed quantum computation and was making strides. It was 2016, when the company announced the creation of a $ 5 $ qubit quantum computer, where one qubit would correct errors, with the computing device being based on a "five-qubit superconducting chip with star geometry and implementation of the complete Clifford algebra\footnote{Algebra that is based on a vector space and is quadratic in form. \cite{Clifford1871,Artin1988}}." \cite{Savchuk2019} The machine was programmable; it allowed for the creation of gates and the modeling of operations. \cite{Savchuk2019} But the progress has not stopped there, as in 2017, in May, to be exact, another announcement was to be made: quantum computers with $ 16 $ and $ 17 $ qubits have been implemented; and then an enormous leap, in November of 2017, IBM announced a quantum device with $ 50 $ qubits, where $ 20 $ qubits were used for computation and $ 30 $ were used for error correction. \cite{Savchuk2019} It was possible for this quantum device to maintain its qubits in a coherent state for up to $ 90 $ $\mu s$, and the device was with consumption of $ 10-15 $ $ kW$ of power "sufficiently energy-efficient"--without including the energy for device cooling outside work. \cite{Savchuk2019}

Quantum computing research was now beyond its fledgling days, and in 2016, the first quantum satellite was launched from China\footnote{A joint project of the Chinese Academy of Sciences (CAS), University of Science and Technology of China, Austrian Academy of Sciences (AAS), and University of Vienna. \cite{AAS2016}}, Micius\footnote{"?470--?391 BC, Chinese religious philosopher; his teaching, expounded in the book Mo-Zi, emphasizes love, frugality, avoidance of aggressive war, and submission to Heaven." \cite{CollinsDictionaries2023}} it was called. \cite{Chen2021} The goal of the space mission was to "perform quantum experiments at space scale", which was an important achievement for quantum communication and space science at the same time. \cite{Chen2021} This attempt at a space-scale quantum leap in 2020 resulted in a new milestone for space quantum communications when, via Micius, a secure link, by quantum key distribution\footnote{In the late 1960s \cite{Nagy2006}, the birth of quantum cryptography occurred with Stephen Wiesner's idea of using quantum mechanics \cite{Savchuk2019}, published in 1983 \cite{Wiesner1983}, in order to produce unforgeable money \cite{Nagy2006}. Even though unpractical, the idea quickened others, like it did Bennett et al. \cite{Bennett2014} who developed the BB84 protocol (as it was originally published in 1984: \url{https://ars.els-cdn.com/content/image/1-s2.0-S0304397514004241-mmc1.pdf}) for quantum key distribution, where secret keys were exchanged securely over a public channel, in contrast to cryptography based on public keys that is so widely used today, here security is achieved by laws of physics that are not in eavesdroppers favor. \cite{Nagy2006,Bennett2014,Savchuk2019}}, was established between two on-ground stations that were separated by 1120 kilometers. \cite{Yin2020} While these events were happening, another breakthrough was in the making.

Intel was interested in quantum computation, and this they loudly expounded in January 2018 when a declaration was made of superconducting quantum chip implementation, the name was Tangle Lake, quite an Intelish name, I might add, and the number of qubits was $ 49 $. \cite{Savchuk2019,Hsu2018} This event was followed by one coming from Google, for they presented in March 2018 a new quantum superconducting processor, Bristlecone, with a capacity of $ 72 $ qubits. \cite{Savchuk2019,Kelly2018} This device was a continuation of a previous one announced a few years ago with $ 9 $ qubits and a rather low level of error, which was $ 1 \% $ for data reading, with $ 0.1 \% $ and $ 0.6\% $ for one-qubit and two-qubit quantum gates, respectively. \cite{Savchuk2019} With a two-dimensional structure of two $ 6 \cdot 6 $ arrays that are placed one above the other, the system can track the errors happening during computation and correct them\footnote{It has been demonstrated by Google's researchers that it would take only $ 49 $ qubits for a quantum advantage, superiority, to happen "if the number of gates exceeds $ 40 $ and the error of two-qubit quantum gates is less than $ 0.5 \% $". \cite{Savchuk2019} A Superiority being, performing a task by quantum computer exponentially faster, super-polynomial speedup is a must here, than on a classical computer; this task can be any task, even a practically useless one. \cite{Preskill2018,Zhong2020,Yin2020}}. \cite{Savchuk2019}

With the ever-moving advance of quantum devices, research was continuing in different aspects of quantum mechanics, an important element for quantum computation, and although evidence is still not conclusive, in 2018, quantum entanglement was observed in objects almost visible to the naked eye, a potential application of which could be seen in quantum internet and physics research. \cite{Riedinger2018,Popkin2018}

In 2019, the Google AI Quantum group declaimed \cite{Preskill2021} "a 52-qubit superconducting chip named Sycamore, which they claim has demonstrated quantum supremacy" \cite{Resch2019,Murgia2019}. A first claim of this type and a very exciting one, however, when one looks back from a distance, only then it is often the case that a man can clearly see what was the event that made something of something; it might be that it was this one, but perhaps it was not just yet. \cite{Savchuk2019,Pednault2019}

As it seems that the previous question has not been answered yet, let us jump to one that is, namely, quantum entanglement. In 2022, after decades of effort and research, it seems that Einstein's "spooky action at a distance" has finally been thoroughly investigated and brought into the realm of fact, since in the year mentioned, Aspect, Clauser, and Zeilinger received the Nobel Prize in Physics \cite{NobelPhysics2022}, and while this research article is not about rewards, a question that has for many decades puzzled some of the best minds deserves a mention\footnote{Alain Aspect and John F. Clauser have contributed to expounding and demonstrating the true nature of quantum entanglement, while Anton Zeilinger linked entangled particles and propagated correlation with such entangled systems, making quantum network. \cite{Billings2022,NobelPhysics2022}}. The last loophole\footnote{There is one more loophole, namely super-determinism, "identified by Bell himself: the possibility that hidden variables could somehow manipulate the experimenters’ choices of what properties to measure, tricking them into thinking quantum theory is correct." \cite{Merali2015}} in a well known Bell's test has been closed in 2015 \cite{Hensen2015}, thus supporting quantum theory \cite{Merali2015}, the universe we live in is not anymore strange; it is quantum entangled and magnificently fascinating.

If the previous event seemed imposing, the next one is in at least the same category, for in 2023 a reversing protocol for a quantum system has been demonstrated, with authors noting that this new understanding of quantum mechanics could have application in quantum information technology. \cite{Schiansky2023,Trillo2023} The protocol is a universal time-reversing mechanism with an arbitrarily high probability of success, where interference of different paths in the end causes the system to jump to the state it had some $ T $ time units before--the protocol is "requiring no knowledge of the quantum process to be rewound, is optimal in its running time, and brings quantum rewinding into a regime of practical relevance." \cite{Schiansky2023,Trillo2023}

What brings us at the cusp of time, it is still unknown what technology will prevail \cite{Resch2019}, or if it will perhaps be a mixture of the two, quantum and classical working in tandem, a most probable outcome, but what can be stated with greater certainty is that the next ten years will probably reveal and answer far more in terms of quantum machines usefulness and area of their specialty.

While the history of quantum computing is for the time being concluded, what comes next is an open question, a choice, and a work that is yours.

\section{Quantumness of Quantum Computing}
\label{sec:fQC}
Even though quantum computing has seen great progress, it seems that it is a subject with which scientists and practitioners are still not that familiar. There are probably at least these reasons behind it: their education has not covered the topic, they still do not see the use of such a tool, the state of quantum computing is still far from mainstream, and the link between quantum physics and computing is not an easy one to make. It is also a matter of fact that quantum computation is a sub-discipline that is multidisciplinary in its essence and requires experts with vastly different backgrounds \cite{Nagy2006}, as such, it represents a sub-discipline for which it is difficult to get your brain around.

If one searches through existing scientific papers, a substantial number of articles will now be found, and the articles range from theory to practice, from the synthesis of knowledge to algorithms. Naturally, the papers include important elements that one needs when dealing with quantum computing. It is, however, problematic that so many of these elements present a stumbling block in that learning curve towards quantum way of conducting work--quantum computation is so vastly different from classical computing, and it is perhaps in the beginning quite daunting to come from classical computation, where one knows much, to quantum computation, where one knows little.

For example, right at the start of one's journey to the universe of quantum, instead of a classical bit for information storage, one is confronted with a quantum bit, i.e. a qubit. And instead of storing one value, as in a bit, in quantum computation, one has a situation where one qubit is in both states \cite{Mooij1999} (both levels) simultaneously.\footnote{Quantum computer can also be a three-state system as a qutrit \cite{Xiao2013}, or can even be in a more complex multi-level, $ d>2 $, state as a qudit \cite{Wang2020}, with a number of qubits in a group being denoted as a register.}

After dealing with the qubit question, one is confronted with other quantum computing peculiarities like entanglement and collapse of quantum states through observation. It is almost one big thing after another, and to get to grips with these and other questions, the goals of this research article on quantum computation are the following:

\begin{description}
	\item [Historical Outlook] Develop a never-before-published historical context of quantum computing that is encompassing and detailed without missing valuable information, precise, covering milestones, and presenting the most significant achievements.
	
	\item [Theory Chronology] Synthesize a one-of-a-kind broad, deep, precise, and thoroughly referenced chronological outlook on quantum computing, both textually and visually, through a timeline presenting a broad picture of the field and segments of its history that will expound on the progression of the theory, present those that came before, and show links between quantum phenomena and other fields.
	
	\item [Foundational Terminology] The basis of any theory, together with axioms, theorems, lemmas, and corollaries, is its terminology and definitions of those terms. The literature at the moment offers no complete, deep, and well-referenced material. Such a state of the matter leads to confusion and a lack of understanding in terms of quantum computing. A compendium of such nature is therefore a must; thus, to construct and present such a work is one of the goals of this paper.
	
	\item [Standard Model] As a way of delving into the practical part of quantum computing and gearing toward computer experts in a streamlined and straightforward manner\footnote{Even though the focus group of this paper is computing experts, the review is written in such a way that anybody with basic computing and mathematical knowledge should be able to understand it.}, quantum computing knowledge will be combined through a standard model\footnote{The circuit model of quantum computing \cite{Deutsch1989}, which is the most amenable from an algorithmic perspective [66], consists of a sequence of quantum gates (unitary operations). "Thus, quantum languages and compilers should facilitate the conversion from high-level descriptions to individual gates and the control signals necessary to perform them." \cite{Resch2019}} of computation, with special emphasis on foundational high-level quantum algorithm modeling and a design pattern.
	
	\item [General Outlook] Synthesis of the present state of the art with the future importance and possibilities of quantum computing. Embedding discussion on problems still in need of solving while not forgetting those pervasive open questions.
	
	\item [From Now to Beyond] Provide a number of quality literature materials that will present themselves as an extended arm of this research. Facilitating an even broader reach of the research conducted and enabling future research and algorithm development through a compact number of reliable steps to the next breakthroughs and game-changers.
\end{description}

In order to achieve previous goals, an effort will be made to cater to the computer science mind and to build a strong theoretical foundation and intuition. Thus enabling a correct, consistent, and deep understanding of quantum computation and quantum mechanics' phenomena. With the introduction over, the next step in the journey is foundational terminology.

\section{Foundational Terminology}
\label{sec:Terminology}
When one is dealing with any subject, there are primarily two ways in which he can proceed to expose the issue. The first is to start with a general and then build in a top-down manner. The second is, of course, to start with concrete and then build in a bottom-up manner. They both have their pros and cons, with the latter being more fascinating and interesting, but perhaps in certain instances it is more difficult to understand in such a way, with the former being more conceptual and gradual, but not a stumbling block on the mind while trying to grasp some complex new idea. One would choose one or the other depending on the subject, audience, and perhaps some other factors as well.

It is often the case, perhaps even exclusively, in the scientific literature, at least in the discipline of quantum computation, that the more practical approach, which is bottom-up, is used. Considering that quantum computation at its best is physics in action, that approach is logical and has its merits. However, quantum mechanics is so strange and at times so counter-intuitive that it is quite challenging to understand its complex essence, and the mind has an issue combing all those different threads of thought at the same time--for thinking, one needs time, and for thinking about quantum computation, one needs a considerable amount of time. And if learning is impeded, if the subject has not been understood, one cannot expect great results from then on.

Therefore, in order to continue the strain of thought from previous sections, to give the mind the necessary time for information incubation, and to build up essential intuition, before we delve into some concrete examples of quantum computation essential for the review and an outlook that is being written, we will first define a broad range of terms\footnote{The terms defined will in some instances perhaps be of a broader interest than this paper would require, nevertheless, to leave no stone unturned and to give a comprehensive review of foundational quantum computation terminology, this will be done.} that will be linked to that practical quantum computation and revealing of fascinating knowledge about it, but not so overwhelming that it will impede progress more than it would be expected. The first stop will then, fittingly, be the definition of quantum mechanics.

\textit{Quantum Mechanics} It is said of physicists that quantum mechanics represents the most complete as well as the most accurate description of the universe we live in. \cite{Nagy2006} It is a theory consisting of rules and principles that define a framework that is then, in turn, used in order to develop other physical theories. \cite{Nagy2006} What these rules, principles, and mathematics are, we will soon see.

\textit{Quantum Computing} The act of using those rules and principles of quantum mechanics in order to carry out computation is then called quantum computing. \cite{Resch2019} Quantum computing has two powerful mechanisms through which computation is performed, namely superposition and entanglement, and these have no counterpart in classical computation. \cite{Resch2019} Such is the nature of computation that is quantum, and these are its key advantages. \cite{Resch2019} It is well known what data is and what information is, but how is that transferred into the realm of quantum? We will answer that next.

\textit{Quantum Information\footnote{Information and data are often used interchangeably, although there is a difference. Data represents a fact about the world we live in, while information represents newness extracted from data, which then becomes data as well.}} Those well-established definitions and understandings of data and information are at a general level unchanged; however, at the practical level, the situation is quite different. According to the well-known no-cloning theorem, quantum data cannot be copied, and as such, it lasts only as long as the program lasts\footnote{The superposition of a qubit when observed collapses, and there is no way to multiplicatively transfer or amplify a quantum state so as to admit a number of copies of a quantum system. \cite{Wootters1982}}. \cite{Wootters1982,Resch2019} Data is, to a physicist, an encodable and storable feature that can be processed "in some physical system using some physical process." \cite{Preskill2021} Data may then be regarded as a feature that one stores and processes in a quantum state. \cite{Preskill2021}

\textit{Quantum Bit\footnote{Termed by Benjamin Schumacher. \cite{Schumacher1995}}} A qubit, or quantum bit, represents an indivisible unit of quantum data. \cite{Preskill2021} Abstract qubits can be encoded in a physical quantum system, and that qubit can be "an atom, an electron, a photon, an electrical circuit, or something else." \cite{Preskill2021} Unlike a classical bit that can be $ 0 $ or $ 1 $, a qubit can be in multiple states simultaneously, mathematically described as a vector in a complex Hilbert space\footnote{Real or complex vector space that is higher dimensional, may be infinite, generalizes linear algebra and calculus, sequences of which are convergent, and provides a distance function. \cite{Debnath2005,Carlson2023}}, "with two mutually orthogonal basis states which we can label $ |0\rangle $ and $ |1\rangle $." \cite{Preskill2021} These orthonormal states can, for example, correspond to a different polarization of a photon or perhaps to a different spin of an electron. \cite{Rieffel2000}

\begin{figure}[h]% superposition
	\centering
	\includegraphics[width=0.9\textwidth]{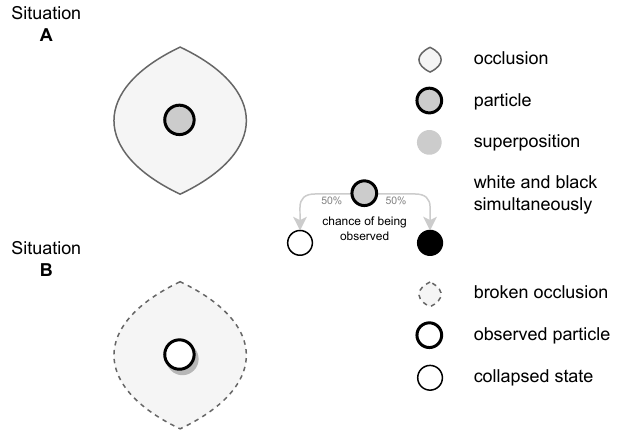}
	\caption{Illustrative example of a quantum phenomenon known as superposition. Under, for example, a measurement, superposition would collapse, and one would observe either a white or a black state, or a white or black ball in this instance.}\label{fig:sp}
\end{figure}

\textit{Superposition} Feature of being quantized, Fig. \ref{fig:sp}, and having infinite degrees of freedom, that is, being in multiple states\footnote{Quantum state can be pure (represented via state vector and not mixed with other states) or mixed otherwise (represented via density matrix and a mixture of states). \cite{Mintert2007,Pusey2012}} at the same time (linear combination)--until observation has been made. \cite{Swenson1972,Zeh1970} This feature represents one of the two main pillars of quantum mechanics, the other being entanglement. \cite{Bouwmeester1999} Through superposition, one has access to the real power of quantum computation via the exponential state space of multiple qubits. \cite{Rieffel2000} "Just as a single qubit can be in a superposition of $ 0 $ and $ 1 $, a register of $ n $ qubits can be in a superposition of all $ 2^n $ possible values." \cite{Rieffel2000}

\begin{figure}[h]% entanglement
	\centering
	\includegraphics[width=0.9\textwidth]{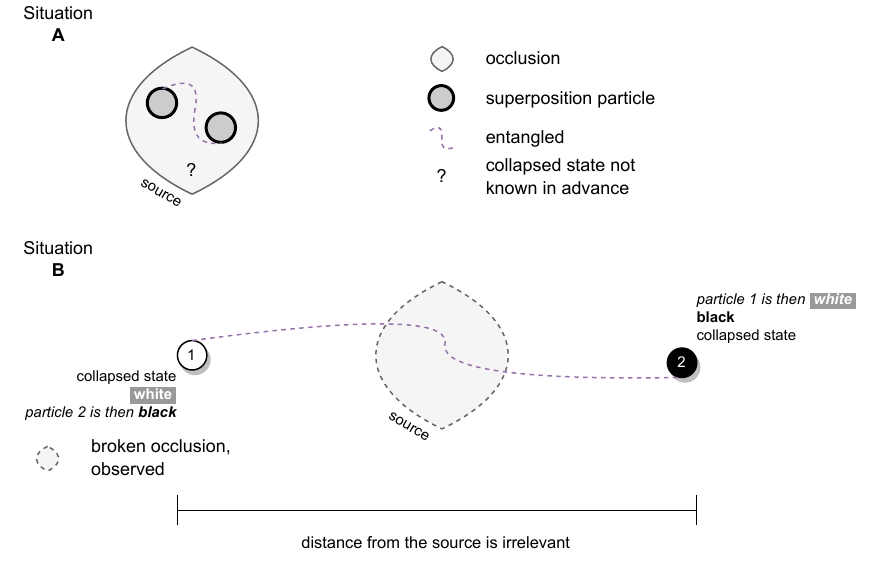}
	\caption{Illustrative example of a quantum phenomenon known as entanglement. While particles are entangled, and as experiments have shown there is no hidden variable involved, they are influencing one another to such a degree that either party can predict the state in which a particle of the other side is when observed, no matter the distance between parties.}\label{fig:ent}
\end{figure}

\textit{Entanglement} Quantum state where particles, Fig. \ref{fig:ent}, and in quantum computing qubits, are locked, with one exhibiting an influence on the other (there is a correlation between particle states, e.g. one particle collapses to $ 0 $, the consequence of which is that the other then measures to $ 1 $). \cite{Resch2019} Distance between particles does not play a role; that is, entanglement correlation works regardless of the distance\footnote{In order to entangle particles, they need to be brought close together so as to interact, and then they can be sent long distances. \cite{Preskill2021} With today's technological limitations, it is challenging to send an entangled qubit very far, i.e. from Pasadena to New York, without damaging the qubit state during travel. \cite{Preskill2021}} between particles--this is a phenomenon of which Einstein did not speak so kindly when he said, "spooky action at a distance" \cite{Resch2019}, but it turned out to be correct nevertheless \cite{NPrize2022,Pan2000}. Data is in quantum computation and is therefore stored both in qubits and in relationships between them, with the amount of stored data being exponential in the number of qubits. \cite{Preskill2021}

\textit{Bell State} Quantum state, also known as EPR\footnote{The reader should take note that the well-known EPR paradox deals with incompleteness, while Bell's theorem deals with the non-locality of correlations. \cite{Einstein1935,Bell1964}} (Einstein, Podolsky, Rosen) pair of two qubits that are in superposition and are maximally, in regard to correlation, quantumly entangled. \cite{Sych2009,Bell1964} These Bell states can be both symmetric and asymmetric (e.g. $ 1 $ and $ 1 $, or $ 1 $ and $ 0 $), \cite{Sych2009,Galindo2022} with applications in quantum teleportation \cite{Sheng2010}, dense coding \cite{Piveteau2022}, information processing \cite{Deng2017}, privacy protection \cite{Lang2021}, cryptography \cite{Thapliyal2015}, networks \cite{Zhang2016}, optics \cite{Leonhardt1995}, etc.

\textit{Teleportation} Enabled by particles that are in a quantum state and entangled, where an unknown particle state is transferred between far apart parties, from one party to another, from one particle to another, but the particle itself is not sent. \cite{AHARONOV1999} In the procedure for such an event, before teleportation can take place, some source $ S $ needs to generate an entangled pair and send particles to their respective destinations. \cite{Pirandola2015} Then, when quantum communication can begin and data transfer happen, after one side has made a measurement, the other side needs to be contacted via classical channels, bound by no faster than light communication, so as to inform them of the measurement parameters for observation, through which the other side will ultimately receive quantum data via the obtained state. \cite{Rieffel2000,Pirandola2015}

\begin{figure}[h]% dense coding
	\centering
	\includegraphics[width=0.9\textwidth]{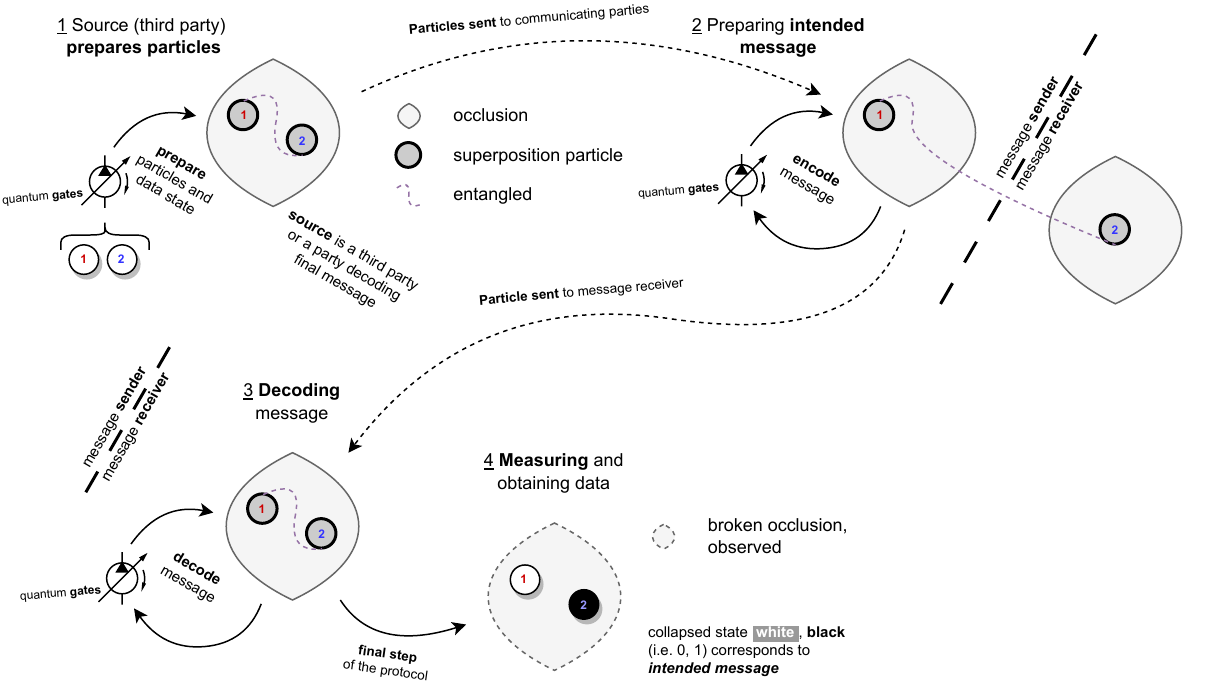}
	\caption{Illustrative example of a quantum phenomenon known as dense coding. The receiver would, in order to decode the sent message, employ a series of quantum gates (C-NOT and afterword Hadamard--more of which will be discussed in the continuation), but in the reverse order of the source party that has, in order to put particles into superposition and an entangled state with appropriate data, prepared particles for transit. Thus, through a series of steps, particles have been prepared and moved to parties involved in communication, with the sender of the message encoding the message through a received particle and sending the particle furthermore to the receiver of the message, who, at the end of the line, decodes and reads the message intended for him. The symbol for quantum gates via lines connected to a circle indicates input and output; the circle itself represents an enclosure that holds a superposition of states represented by both white and black surroundings; and two lines represent changes of different variables/characteristics.}\label{fig:denc}
\end{figure}

\textit{Dense Coding\footnote{Also known as super-dense coding.}} Protocol that is dual to teleportation, Fig. \ref{fig:denc}, and depends on the entanglement that is described in the EPR experiment; it uses a single qubit in order to transfer, that is, transmit, two bits (in terms of classical information). \cite{Rieffel2000} If source and destination have a particle of EPR entangled pair with maximal correlation, which source has prepared and then sent one particle to destination, it is then possible to transmit two bits of classical data via only one qubit by applying a unitary operator at destination and returning that particle back to source, where party at the source can now jointly measure both particles, that is, the entire EPR pair, and naturally, also learn of the operator party at the destination used in order to manipulate the particle that it received. \cite{Bennett1992}

\begin{figure}[h]% measuring
	\centering
	\includegraphics[width=1\textwidth]{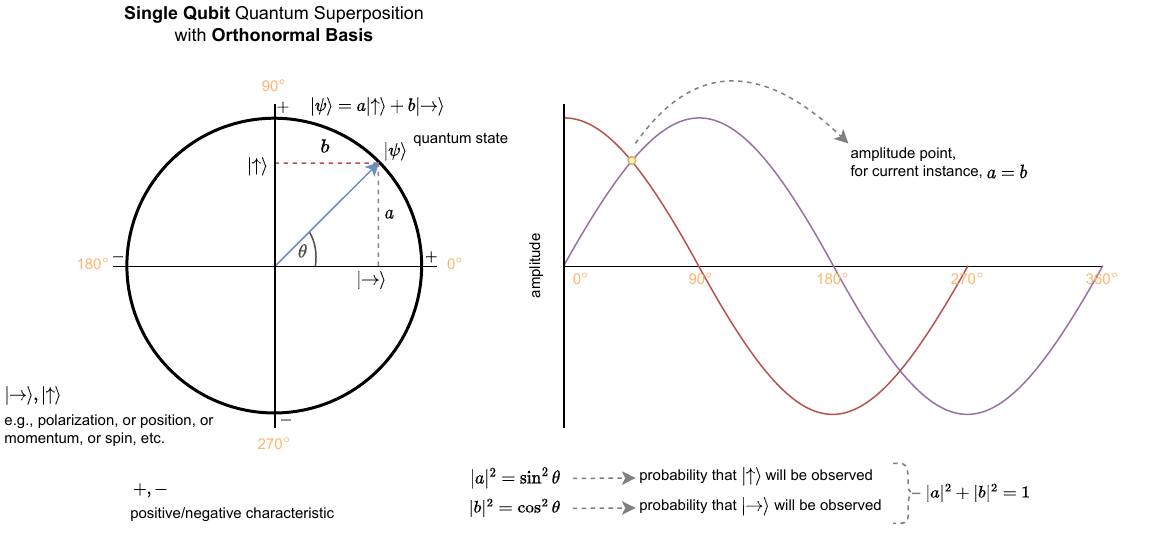}
	\caption{Illustrative example of a one-qubit quantum measurement. \cite{Rieffel2000} An enclosure of one qubit is presented, with no algorithmic influence on that qubit. Qubit is in a superposition of two orthonormal states: $ \ket{\uparrow} = a$ and $ \ket{\rightarrow} = b $--which means that quantum superposition state is $ \ket{\psi} = a\ket{\uparrow} + b\ket{\rightarrow} $. $ \ket{expression} $ is a part of the braket notation (more of which will be expounded further on), which is used to express quantum states. As the amplitude point is defined by a $ 45^\circ$ angle $\theta$, both $ a $ and $ b $ are equal, therefore $ a = b = \frac{\sqrt{2}}{2} = \frac{1}{\sqrt{2}} $. Since we are dealing with orthogonal unit vectors, amplitude values can be normalized into state probabilities as $ |a|^2+|b|^2=1 = \left(\frac{1}{\sqrt{2}}\right)^2 + \left(\frac{1}{\sqrt{2}}\right)^2$, which corresponds to the total probability of the system, with $ |a|^2 $ equaling probability for $ \ket{\uparrow} $ and $ |b|^2 $ equaling probability for $ \ket{\rightarrow} $--known also as the Born rule \cite{Landsman2009}. After measurement, $\ket{\psi}$ will collapse to $ \ket{\uparrow} $ with a $ 50 \% $ chance, and any subsequent measurement of the same basis will yield the same measured state with a probability of $ 1 $--the original state is lost and it is not possible to determine what it was. \cite{Rieffel2000} The example deliberately uses an instance with amplitudes resulting in equal probabilities; take note that this is for illustrative purposes only, as amplitudes and consequently probabilities vary depending on initial state preparation, quantum circuit, etc.}\label{fig:meas}
\end{figure}

\textit{Measurement} Disturbing the quantum state by making an observation, Fig. \ref{fig:meas}, intended or otherwise. \cite{Rieffel2000} Quantum measurement\footnote{In order to represent a qubit in three-dimensional space, one would use a Bloch sphere \cite{Wie2020,Kashmadze2017}, named after Nobel Prize winner Felix Bloch \cite{NPrize1952}, useful for representing together quantum gates, observations, as well as quantum states.} is probabilistic, and it is not an easy task to "pick" the result one would like to receive. \cite{Rieffel2000} Since data from a qubit can only be obtained by measurement, regardless of the superposition of states, in the end it is possible to extract only one classical state, in terms of data, from a qubit--and the reason is that when measurement takes place, the superposition collapses and the state is changed to one of the basis states. \cite{Rieffel2000} In order to describe the phenomena of quantum mechanics, scientists have used complex numbers, but as the imaginary part of the phenomenon description is not observable in the physical world, out of the four dimensions that we would need for two base states of quantum computing, one would have only two dimensions; thus, the Bloch sphere has three dimensions, two for polarization and one for the base states. \cite{Sutor2019} New information is however coming into focus, as it seems that there are entangled states that are distinguishable only by their imaginary component \cite{Wu2021,Wu2021a,Renou2021,Chen2022,Li2022}--as fascinating as these discoveries are, whether the imaginary number mathematical trick used to facilitate calculations is necessary for the physical world is yet to be determined via the mountain of evidence that future research needs to provide.

\textit{Quantum Gate} An operator, also known as a quantum logic gate, is used to both create and manipulate quantum states. \cite{Monroe1995,Zhou2000} It is an elementary quantum circuit that makes operations on a small number of quantum bits. \cite{Monroe1995,Zhou2000} With these, one is building a complex quantum circuit, and this complex circuit is enabling the execution of an algorithm on the quantum machine. \cite{Monroe1995,Zhou2000}

\textit{Quantum Circuit} Model of computation consisting of a series of qubits (or some sort of quantum data storage), initializations, gates, and measurements. \cite{DiVincenzo2000,Chiribella2008,Fisher2023}

\textit{Quantum Algorithm\footnote{Take note that classical algorithms can be run on a quantum computer, and at times qunatum algorithms use certain classical algorithms. \cite{Shor1994}}} An algorithm, much like a classical algorithm, that uses quantum effects and represents a sequence of steps, which in turn, by a number of operations, manipulate the initial quantum state for some input, and at the final stage, with measurement being taken, the algorithm returns the correct answer. \cite{AHARONOV1999}

\textit{Quantum Parallelism} The effect present in quantum systems where the amount of parallelism increases exponentially as the size of the system itself, that is, the physical space required, increases linearly. \cite{Deutsch1992} As $ n $ qubits allow one to work at the same time with $ 2^n $ states, quantum parallelism is the effect that gives quantum computing its superiority as it bypasses the classical restriction of time/space tradeoff by giving an exponential quantity of computation space in a linear quantity of real physical space, and therefore quantum machines can compute solutions to all possibilities at the same time, while classical computers can compute only for one input state at the same time. \cite{Rieffel2000}

\begin{figure}[h]% interference
	\centering
	\includegraphics[width=0.9\textwidth]{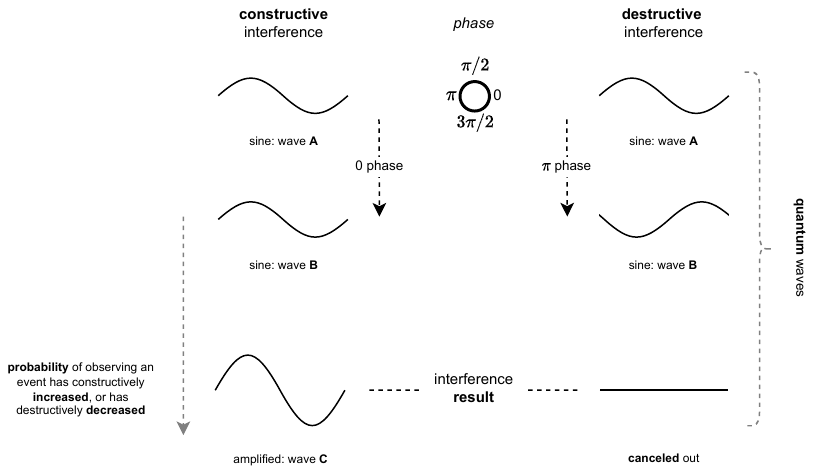}
	\caption{Illustrative example of a quantum phenomena known as interference. As tiny particles behave like waves, those waves interfere and either amplify or inhibit each other.}\label{fig:intrfr}
\end{figure}

\textit{Interference} When measurement is performed on a superposition of output states for a particular input, what one will receive is a random collapse to one state out of all states in the superposition, Fig. \ref{fig:intrfr}, with all other states, that is, values, being destroyed. \cite{Rieffel2000} In this way, one cannot reliably compute, and such a behavior needs to somehow be guided. Interference allows us to do exactly that, guide towards desirable output. With interference, it is possible to cause a cancellation between exponentially many input parallel states\footnote{As an example, one can think about waves of the sea that are interfering one with another, or rays of light. \cite{AHARONOV1999}}, with the goal being to produce such an interference between states, that is of the wave function\footnote{Description of a quantum state through amplitudes and probabilities that can be derived from those amplitudes, typically referred to as $\psi$ or $\Psi$. \cite{Raymer1997,Goldstein2011}}, so as to destroy all undesirable states and collapse into exactly the one we need. \cite{AHARONOV1999} The combination of quantum parallelism and interference gives quantum computation tremendous power, and its use in quantum algorithms is essential. \cite{AHARONOV1999,Chen2021}

\begin{figure}[h]% decoherence
	\centering
	\includegraphics[width=0.9\textwidth]{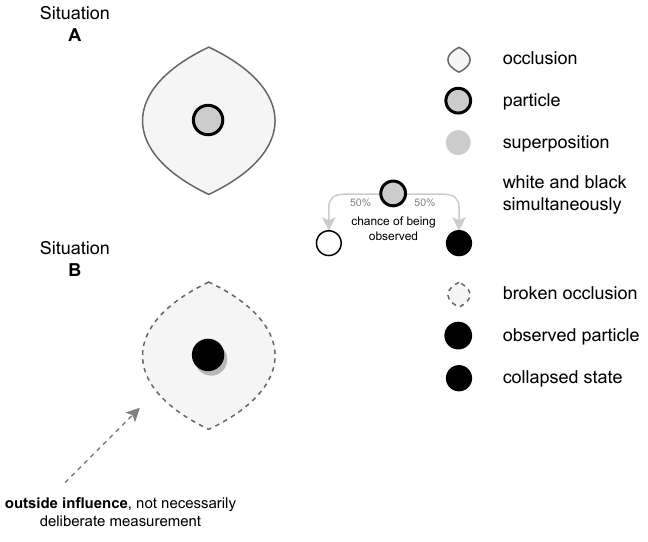}
	\caption{Illustrative example of a quantum phenomenon known as decoherence. Under outside influence on a particle and interaction with the surrounding system, the state of the particle will collapse and its superposition will be destroyed; the effect is the same as when one would deliberately measure a quantum system; this effect is called decoherence. Isolation of particles is typically done via vacuum or cooling to an almost absolute zero ($ -273.15^\circ C $, which is $ 0 K$) temperature--the more a particle is isolated, the easier it is to control it and for it to stay in a superposition for a longer period of time. \cite{Gerrits2010,Sinha1997}}\label{fig:dechrc}
\end{figure}

\textit{Decoherence} For the reason of the interaction of the quantum system and its environment, which is inevitable, the state of a quantum system is extremely fragile, Fig. \ref{fig:dechrc}, and thus due to this interaction, the quantum nature of the system can be lost--this loss of quantum information, this distortion\footnote{Decoherence is the most difficult problem to tackle in quantum computation, as it is extremely difficult to isolate a quantum system from its environment, and it was feared that for this reason alone a quantum computer could not be built, but through the invention of quantum error correcting codes, this stepping stone was overcome. \cite{Rieffel2000}}, and collapse of superposition due to interaction of the quantum system with its surroundings is called decoherence. \cite{AHARONOV1999,Rieffel2000}

By reading the text to this point, a first quantum computation has already been performed; in fact, probably more than a few were done in one's mind. This incubation of data and information has not only made one knowledgeable about the subject of quantum computing but has also developed intuition and a crucial way of thinking needed for such a topic as quantum computation. And now, with neurons and pathways of the brain speaking quantum computation, we will deal in a bit more detail with topics that were touched upon, but for which one's scientific curiosity, trying to decode the universe we live in, wants more.

\section{Quantum Effects and the Universe we Live in}
\label{sec:QEP}

Some time has passed since the event, when during one of his talks, Nobel Prize winner Ivar Giaever told the story of his youth and a job that he applied for and received. After getting the job, his mentor told him a story about quantum mechanics, more specifically tunneling--the story was so strange that Giaever's own words will best explain his disposition: "I did not believe a word of what he told me, nothing." Giaever got his job and decided to be quiet, but what is it that his mentor, John Fisher, told young Giaever?

He told him a story of small particles; he told him a story of the underlying laws of physics that are the foundation of the world we live in; he told him that if one would throw a tennis ball in the wall, that ball would eventually cross the wall and end up on the other side, in the same condition in which it was before it went through the wall; and to top it off, he told Giaever that there would be no hole in the wall. Now that was some story, like something from a fairytale, and Giaever's reaction of not believing a word of what he was told was expected.

During that fascinating talk, Giaever expositioned, explaining that what if one would take an extremely small particle for a ball, i.e. electron, and throw that ball toward obstacles that are very close to one another, distanced in a few atoms, and are not touching? In that case, there is a finite probability that an electron will find itself on the other side of the obstacle, never being in between. That is quantum tunneling, and that is for what Giaever shared his Nobel Prize, in 1973. \cite{NPrize1973} Quantum effects are real and are typically observable only with very small particles\footnote{There is an effect known as the Josephson effect where quantum phenomena are observed on a macroscopic level. \cite{Josephson1962} It occurs when insulating material is placed in between two superconductors, resulting in a tunneling super-current that flows across the junction. \cite{Josephson1974}}, on an atomic and subatomic level.

\begin{figure}[h]% tunneling
	\centering
	\includegraphics[width=1\textwidth]{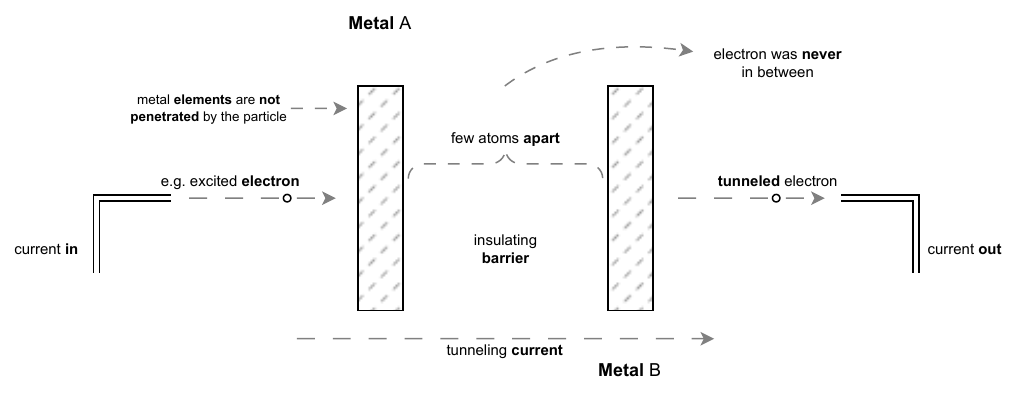}
	\caption{Illustrative example of a quantum tunneling effect. By constructing an electric circuit with two metal elements being a part of that circuit, there is a finite chance that an insulating barrier in between those metal elements will not stop the current if the elements are extremely close to each other (the barrier being no thicker than 100 angstroms); this current is called a tunneling current and is a consequence of wave-particle duality and quantum probability amplitudes of many electrons "attacking" the insulating barrier--metal elements will not be penetrated and, e.g. electrons will never occupy the insulating barrier. \cite{Giaever1974} For this reason, there are limitations/issues with technology miniaturization, as electrons in nanotechnology tunnel through insulating barriers and semiconductor devices. \cite{Li2013,Malinowski2019}}\label{fig:tunn}
\end{figure}

Quantum effects are dependent on a wave function and its accompanying probability that a state will be observed. This wave-particle duality was an outstanding discovery with profound consequences that are shaking science to this day. \cite{Angelo2015} The differences between classical and quantum systems are many, with one of the more intricate being the answer to the question of what one knows about one type of system and, of course, the other. If we know everything about a classical system, including all its characteristics, we naturally know everything about its components; however, this does not hold for quantum systems, which can clearly be seen in a quantum phenomenon called entanglement. \cite{Preskill2021,Pusey2012,Schroedinger1935} One could, for example, have a composite quantum system, i.e. $ AB $, and know everything about that system's laws of physics would allow us to know, despite of that fact, if one would observe just part of the system, i.e. $ B $, information needed to completely characterize that part of the system is missing, as the series of expectations for the subsystem depends on an unknown value of the variable for some other subsystem, in this instance, on the observation of $ A $. \cite{Schroedinger1935,Preskill2021}

This series of expectations, being a consequence of superposition, is linked to entanglement in a meaningful way. Extra states, with no analog in a classical system, leading "to the exponential size of the quantum state space are the entangled
states". \cite{Rieffel2000} In this way, by undergoing initialization, quantum state transformations, and measurement, a quantum system achieves its result. \cite{Rieffel2000}

Even though we live in the quantum world, seldom do we think about it, but the macroscopic world we are surrounded with is not isolated from its own surroundings and is therefore in uninterrupted interactions with the environment, meaning it is continually measured, a phenomenon aforementioned and called decoherence. \cite{Preskill2021} Such a quantum system, continually being observed, represents a system known from the down of time and "is well described by classical physics." \cite{Preskill2021} Though "weird," a vast number of experiments have shown that quantum mechanics correctly describes physical reality. In order to combat decoherence in quantum computers, a breakthrough came, but not from the physical side, as was perhaps expected. \cite{Rieffel2000} It was theorized by some that quantum error correction is beyond our abilities "because of the impossibility of reliably copying an unknown quantum state", yet it was not so, as it is possible via error-correcting techniques to design error-correcting codes by which one can detect specific errors and reconstruct "the exact error-free quantum state." \cite{Rieffel2000}

And so this battle between decoherence and superposition continually "rages." A quantum system can perform an enormous amount of computation in parallel, but accessing the desired result is far from easy. \cite{Rieffel2000,Nagy2006} In order to read the result, quantum state is disturbed, only one of those parallel threads is read, and as the measurement is probabilistic in nature, "we cannot even choose which one we get." \cite{Rieffel2000} It is, however, possible to skillfully deal with the problem of measurement and thus exploit quantum parallelism; "this sort of manipulation has no classical analog and requires nontraditional programming techniques." \cite{Rieffel2000} Shor’s factorization algorithm manipulates quantum states in such a way that the "common property of all of the output values can be read off" \cite{Rieffel2000}, and in this way direct toward the output one would like to achieve, while, for example, Grover’s search algorithm makes amplification through which the probability that the result of interest will be read is increased, thus manipulating quantum states. \cite{Grover1996,Shor1994,Rieffel2000}

Basic operations in any classical algorithm are data copying and data deletion. While trying to project this to a quantum computer, one comes to a brick wall, as this is not possible in a quantum system, perfect copying of an unknown quantum state is an intrinsic impossibility, as per the no-cloning theorem\footnote{Theorem stating, "No quantum operation exists that can duplicate perfectly an arbitrary quantum state." \cite{Scarani2005}}, not just a limitation of laboratory conditions. \cite{Rieffel2000,Scarani2005} If, on the other hand, we disregard the notion that the copy needs to be perfect (producing a perfect copy of a limited number of quantum states with probability $ <1 $), then one can devise an apparatus (a copier or cloner) by which copying can be conducted, reproducing the desired state through an approximation or to a degree of probability. \cite{Hillery2000}

As one might presume, with copying being such a stepping stone, data deletion also differs substantially from the classical case. If one assumes that there are two identical copies of an arbitrary and unknown quantum state to be deleted, this process actually cannot be accomplished (as per the no-deleting theorem), aside from deleting approximately\footnote{It has also been proven that quantum information cannot be split into complementary parts, which demonstrates that an unknown qubit state represents one entity. \cite{Zhou2006}}--as is the case for quantum cloning; however, just as is the case for cloning, the process of deletion is possible if one deals with known orthogonal states. \cite{Samal2011,Pati2000} This inability to clone and delete quantum information, but only express possibilities already in existence, postulates conservation of quantum information, as information cannot be created nor destroyed. \cite{Pati2000,Zyga2011}

Related to cloning and deleting quantum information is the inability to hide information, known as the no-hiding theorem. \cite{Pati2000} If a quantum system interacts with its surroundings and loses information, that information actually is not missing; it simply resides somewhere else in the universe--that is, correlations between the system and the environment are not able to hide information\footnote{These three: no-cloning, no-deleting, and no-hiding theorems postulate the law of conservation of quantum information; just as the energy of a closed system is conserved, so is the information. \cite{Lie2020}}. \cite{Pati2000,Zyga2011}

In spite of all the hurdles we go through when trying to discover new knowledge and understand the universe in which we are, this same quantum universe works perfectly and mindbogglingly precise, with quantum computers being devised and in operation. That being said, scientific discovery and painstaking experimentation have produced criteria for successful implementation of a device that would be called a quantum computer; they are found in \cite{DiVincenzo2000}, and are as follows:

\renewcommand{\labelenumi}{\Roman{enumi}}

\begin{enumerate}
	\item "A scalable physical system with well characterized qubits", that is, a collection of qubits with physical parameters that are accurately known,
	\item "The ability to initialize the state of the qubits to a simple fiducial state, such as $ 000 $", that is, initializing quantum registers to a known value before one starts computing,
	\item "Long relevant decoherence times, much longer than the gate operation time", that is, dynamics with the environment brings about quantum state decay with which quantum computation is possible,
	\item "A 'universal' set\footnote{However, this is impossible, as the number of quantum gates is uncountable; therefore, one requires a finite set of quantum gates that are in a finite sequence of gates approximating any operation. \cite{Sohn2018,DiVincenzo2000}} of quantum gates", that is, a set of quantum gates that are able to implement via a finite sequence of gates any quantum operation,
	\item "A qubit-specific measurement capability", that is, the capacity to be able to measure specific qubits.
\end{enumerate}

In addition to the previous five, two additional ones are added, namely "the ability to inter-convert stationary and flying qubits" and "the ability to faithfully transmit flying qubits between specified locations", in order to achieve quantum communication, as not all information processing is only computation. \cite{DiVincenzo2000} The need for the additional two criteria is clearly seen in quantum key distribution \cite{Bennett2014}, and quantum cryptography \cite{DiVincenzo2000}. It is, however, not an easy task to transmit a qubit from one place to another, and when this is done, decoherence plays an important hurdle to overcome. \cite{Glancy2004,Preskill2021}

In spite of all of its strangeness, quantum mechanics has withstood the test of time, and for the time being, it stands supreme. But just as is the case for the theory of relativity and Newtonian physics, so is the case for quantum and classical physics; both are needed. In fact, classical is quantum, but simply for large objects for which wavelengths are so small that they cannot be measured. Thus, if something functions specifically, it does not mean that it functions generally, but if it does not function generally, it does not mean that it is not useful. With the next section most definitely being useful, as it deals with quantum gates and algorithms.

\section{Computation with Quantum Gates}
\label{sec:QGates}

Fundamentally speaking, as is the case when one does classical computation, by analogy, so is the situation for quantum computation, since in order to manipulate quantum information, one needs quantum gates that are then forming a quantum circuit and consequently a quantum algorithm. There is a myriad of quantum gates, e.g. Identity ({I}), Not ({NOT or {PauliX}}), Controlled Not ({CNOT}), Controlled Controlled Not ({CCNOT or Toffoli}), Swap ({SWAP or S}), Hadamard ({H}), Phase (P), etc. \cite{Menon2014,Sousa2006}, with some being a single qubit gate while others are multiple qubit.

Before we proceed into a more in-depth look at quantum computation, we will first expound on a number of quantum gates, as this knowledge is essential for understanding quantum circuits. Let's start with the quantum gate, whose classical equivalent should be known to every computer expert and physicist: the NOT gate. Let us assume that superposition states, from now on, that we will use shall be $ \ket{0} $ and $ \ket{1} $, with $ \ket{\psi} = a\ket{0} + b\ket{1}$. This basis is called the computational or standard basis and is in three-dimensional space represented by the axes Z, therefore the Z-basis, which is "generally the only basis in which we can make measurements of the system." \cite{Crooks2023}

\textit{NOT} Not gate is a single qubit gate. \cite{Menon2014} Denoted as well as PauliX (named after Wolfgang Pauli, who received the Nobel Prize in Physics in 1945, proposing "that no two electrons in an atom could have identical sets of quantum numbers" that correspond to "distinct states of energy and movement." \cite{NPrize1945}), as the operation it makes is a rotation by $\pi$ radians around the $ X $ axis. \cite{Menon2014} As a consequence of this rotation, there is a mapping, $ \ket{0} \rightarrow \ket{1} $ and $ \ket{1} \rightarrow \ket{0} $. \cite{Menon2014} The transformation matrix used in order to calculate an output for the gate and its input is \cite{Menon2014},

\begin{equation}
	\label{eq:not}
	NOT =
	\begin{bmatrix}
		0 & 1 \\
		1 & 0 
	\end{bmatrix}
\end{equation}

\textit{H} Hadamard gate is a single qubit gate. \cite{Menon2014} Known also as the Walsh-Hadamard gate (named after Jacques Hadamard \cite{Just2022} and Joseph Walsh \cite{Hoyer1997}), the gate makes an operation of superposition--for a basis state, the superposition that is created is equal in probability. \cite{Menon2014,Just2022} Superposition is created by making a rotation of $\pi$ radians around the axis between the $ X $ axis and the $ Z $ axis. \cite{Wen2012} As a consequence of this superposition operation, there is a mapping, $ \ket{0} \rightarrow \frac{\ket{0} + \ket{1}}{\sqrt{2}} $ and $ \ket{1} \rightarrow \frac{\ket{0} - \ket{1}}{\sqrt{2}} $. \cite{Menon2014} If we apply the Hadamard operation twice, a particle is placed into a superposition of states and then returned to its original state. \cite{Shepherd2006} The transformation matrix used in order to calculate an output for the gate and its input is \cite{Menon2014},

\begin{equation}
	\label{eq:hadamard}
	H = \dfrac{1}{\sqrt{2}}
	\begin{bmatrix}
		1 & 1 \\
		1 & -1 
	\end{bmatrix}
\end{equation}

\textit{P} Phase gate is a single qubit gate\footnote{There is a gate known as PauliZ (and PauliX, and PauliY) that rotates the qubit around the $ Z $ axis by $\pi$ radians; this gate is a special case of the Phase gate for $\theta = \pi$. \cite{Menon2014}}. \cite{Menon2014} Known also as the Phase Shift gate, as the gate makes an operation of shifting a qubit's phase with probabilities for the qubit staying unchanged, that is, probabilities for basis states, $ \ket{0} $ and $ \ket{1} $, remain the same. \cite{Menon2014} As the phase is shifted, there is a mapping, $ \ket{0} \rightarrow \ket{0} $ and $ \ket{1} \rightarrow e^{i\theta}\ket{1} $, with $\theta$ being a phase shift and the period being $ 2\pi $. \cite{Menon2014,Fujisawa2004} The transformation matrix used in order to calculate an output for the gate and its input is \cite{Menon2014},

\begin{equation}
	P_\theta = 
	\begin{bmatrix}
		1 & 0 \\
		0 & e^{i\theta} 
	\end{bmatrix}
\end{equation}

The term $ e^{i\theta} $ is a part of the well-known Euler's formula, $ e^{i\theta} = \cos(\theta) + i\sin(\theta)$ (a complex number $ x + yi $ that has magnitude $ 1 $ can be stated via the aforementioned formula)--with the numbers sitting on the unit circle in a complex plane, closing an angle $\theta$ with the axis of the circle. \cite{Stillwell2010}

\textit{I} Identity gate is a single qubit gate. \cite{Sutor2019} This gate does not modify the quantum state in any way--it is typically used in a quantum circuit when we want to show what is happening to a qubit at a certain step or when we want to cause a delay (which the researchers sometimes want to do in order to "calculate measurements of the decoherence of a qubit"). \cite{Sutor2019} The transformation matrix used in order to calculate an output for the gate and its input is the identity matrix \cite{Sutor2019},

\begin{equation}
	I = 
	\begin{bmatrix}
		1 & 0 \\
		0 & 1
	\end{bmatrix}
\end{equation}

\textit{CNOT} Controlled Not gate is a two qubit gate. \cite{Menon2014} This gate is very similar to the Not gate, the difference being that the target qubit is flipped only if the first qubit is in an excited state, that is, there is a mapping, $ \ket{00} \rightarrow \ket{00} $ and $ \ket{01} \rightarrow \ket{01} $ and $ \ket{10} \rightarrow \ket{11} $ and $ \ket{11} \rightarrow \ket{10} $. \cite{Rieffel2000} The transformation matrix used in order to calculate an output for the gate and its input is \cite{Rieffel2000},

\begin{equation}
	CNOT = 
	\begin{bmatrix}
		1 & 0 & 0 & 0 \\
		0 & 1 & 0 & 0 \\
		0 & 0 & 0 & 1 \\
		0 & 0 & 1 & 0 \\
	\end{bmatrix}
\end{equation}

\textit{S} Swap gate is a two-qubit gate. As the name suggests, this gate makes an operation of swapping the values of two qubits; the order of the qubits is not important for this gate. \cite{Menon2014} There is also a version of the Swap gate called the Fredkin gate (a three-qubit gate \cite{Menon2014}), which makes an operation of a controlled swap. \cite{Rieffel2000} The transformation matrix used in order to calculate an output for the Swap gate and its input is \cite{Menon2014},

\begin{equation}
	S = 
	\begin{bmatrix}
		1 & 0 & 0 & 0 \\
		0 & 0 & 1 & 0 \\
		0 & 1 & 0 & 0 \\
		0 & 0 & 0 & 1 \\
	\end{bmatrix}
\end{equation}

\textit{CCNOT} Controlled Controlled Not gate is a three-qubit gate. Similarly to the two-qubit Controlled Not gate, this gate takes two controlled qubits, and depending on the values of these, the value of a third qubit is flipped--that is, iff the first two qubits have a value of 1, then the value of a third qubit is flipped. \cite{Rieffel2000} This gate is also known by the name Toffoli gate. \cite{Rieffel2000} The transformation matrix used in order to calculate an output for the gate and its input is \cite{Sutor2019},

\begin{equation}
	CCNOT = 
	\begin{bmatrix}
		1 & 0 & 0 & 0 & 0 & 0 & 0 & 0 \\
		0 & 1 & 0 & 0 & 0 & 0 & 0 & 0 \\
		0 & 0 & 1 & 0 & 0 & 0 & 0 & 0 \\
		0 & 0 & 0 & 1 & 0 & 0 & 0 & 0 \\
		0 & 0 & 0 & 0 & 1 & 0 & 0 & 0 \\
		0 & 0 & 0 & 0 & 0 & 1 & 0 & 0 \\
		0 & 0 & 0 & 0 & 0 & 0 & 0 & 1 \\
		0 & 0 & 0 & 0 & 0 & 0 & 1 & 0 \\
	\end{bmatrix}
\end{equation}

While not all of these will be used in our own calculations, they represent some of the elementary quantum gates and foundational quantum operations and are therefore mentioned as part of one's necessary quantum arsenal. For a number of other quantum gates, both frequently and infrequently in use, one can consult \cite{Crooks2023}. It is useful to have quantum transformations represented graphically\footnote{By browsing through the quantum literature, one will also find a representation called the Bloch sphere, which is a three-dimensional representation of a qubit's state as a point on the surface of such a sphere. \cite{Menon2014,Radtke2005}}, therefore, a single-bit operations are typically graphically represented by labeled boxes, while multiple-qubit operations are typically represented by circles, marks, and lines--as other authors have dealt with this satisfactorily, we will not repeat it here. \cite{Rieffel2000,Crooks2023}

In order to know the output of a quantum algorithm, we need to be able to calculate that result, just like with a classical algorithm. There is, however, a twist in a quantum situation. Since we are dealing with particle states and quantum operations represented by matrices, we transform input into output by using vector notation for probability amplitudes and then calculate the tensor product for the expression, after which we perform matrix multiplication, which in turn transforms amplitudes, which in turn changes probability density and the end result. Let us therefore perform a few interesting calculations.

If we had a qubit that we wanted to place into a superposition of states, we would use the Hadamard gate, abbreviated as \textit{H}. By following the aforementioned procedure, a qubit needs to be had. Let us therefore define the following qubit, $ \ket{\psi_0} = 1\ket{0}+0\ket{1}$. On this qubit, one now needs to apply the \textit{H} gate, an operation needs to be performed on the operand, so as to achieve the desired result, namely, superposition. By placing the qubit amplitudes into a column vector and using the \textit{H} gate matrix, we will have the following.

\begin{equation}
	\label{eq:hadCalc}
	\dfrac{1}{\sqrt{2}}
	\begin{bmatrix}
		1 & 1 \\
		1 & -1 
	\end{bmatrix}
	\begin{bmatrix}
		1 \\
		0  
	\end{bmatrix}
	= \dfrac{1}{\sqrt{2}} + \dfrac{1}{\sqrt{2}}
\end{equation}

Such a result has given us a qubit in a superposition; thus, by performing the above multiplication, we have $ \ket{\psi_1} = \frac{1}{\sqrt{2}}\ket{0}+\frac{1}{\sqrt{2}}\ket{1} $. Perfect, we have a qubit in a superposition with both states having the same amplitude, and by squaring the state values, we see that both states have a $ 50\% $ chance of being observed after superposition collapse. By applying the \textit{H} gate again, what one can freely try, the original state would again be a reality. It is also good to note here that a qubit is always in a superposition, although for the reason that one of the amplitudes is 0, the original state is often not called such.

With the Hadamard-gained superposition state, we can proceed to another operation. Let us next perform the CNOT operation. We know what the CNOT gate does, and we also know that such a gate is a two-qubit gate. With that in mind, we will define one more qubit, $ \ket{\psi_2} = 1\ket{0}+0\ket{1}$. By placing the qubit amplitudes into a column vector\footnote{This is done by calculating tensor product.}, and using the CNOT gate matrix, we will have the following. 

\begin{equation}
	\begin{bmatrix}
		1 & 0 & 0 & 0 \\
		0 & 1 & 0 & 0 \\
		0 & 0 & 0 & 1 \\
		0 & 0 & 1 & 0 \\
	\end{bmatrix}
	\begin{bmatrix}
		\dfrac{1}{\sqrt{2}} \\
		0 \\
		\dfrac{1}{\sqrt{2}} \\
		0
	\end{bmatrix}
	= \dfrac{1}{\sqrt{2}} + 0 + 0 + \dfrac{1}{\sqrt{2}}
\end{equation}

Therefore, the final state of the operation performed is, $ \ket{\psi_3} = \frac{1}{\sqrt{2}}\ket{00} + 0\ket{01} + 0\ket{10} + \frac{1}{\sqrt{2}}\ket{11} = \frac{1}{\sqrt{2}}\ket{00} + \frac{1}{\sqrt{2}}\ket{11} $. The situation we have here is different than the usual CNOT gate example given above, yet if we think about what has actually happened, this is exactly the result one would expect. We have stated that the CNOT gate will flip the target qubit only if the first qubit is raised, which is 1. Here we had a qubit that served as a control in a superposition, with equal amplitudes, while the target qubit was in a state of $ 1\ket{0} $. Therefore, as the control is in a superposition, if the control is 0, then the second qubit would be the same, while if the control were 1, the target would be raised to 1, which means that the resulting states need to be $ \ket{00} + \ket{11} $, precisely what we have obtained by performing calculation. And as the amplitudes are $ \frac{1}{\sqrt{2}} $, this has "spilled" over to the transformed state $ \ket{\psi_3} $. Two states of the $ \ket{\psi_3} $ whose amplitudes are 0 do not represent a logical outcome, as the tensor product pairs are not in line with the CNOT gate operation.

By observing what has happened with the CNOT gate calculation, one might wonder what else might be in store with various gates and qubit states. We will therefore perform one more operation, and that operation will be Swap, denoted with the \textit{S}. Swap gate is a two-qubit gate that swaps qubit states. This time, let us take the qubit with the state $ \ket{\psi_1} = \frac{1}{\sqrt{2}}\ket{0}+\frac{1}{\sqrt{2}}\ket{1} $ and the qubit with the same state denoted $ \ket{\psi_4} $. Yet again, by placing the qubit amplitudes into a column vector and using the S gate matrix, we will have the following. 

\begin{equation}
	\begin{bmatrix}
		1 & 0 & 0 & 0 \\
		0 & 0 & 1 & 0 \\
		0 & 1 & 0 & 0 \\
		0 & 0 & 0 & 1 \\
	\end{bmatrix}
	\begin{bmatrix}
		\dfrac{1}{2} \\[0.7em]
		\dfrac{1}{2} \\[0.7em]
		\dfrac{1}{2} \\[0.7em]
		\dfrac{1}{2}
	\end{bmatrix}
	= \dfrac{1}{2} + \dfrac{1}{2} + \dfrac{1}{2} + \dfrac{1}{2}
\end{equation}

The final state of the operation performed is, $ \ket{\psi_5} = \frac{1}{2}\ket{00} + \frac{1}{2}\ket{01} + \frac{1}{2}\ket{10} + \frac{1}{2}\ket{11} $. This example of the S gate is perhaps not as intuitive as the one where we have qubits in extreme states, that is, in $ \ket{0} $ and in $ \ket{1} $--with a probability of 1. However, we observe that amplitudes are present for every individual state for both qubits, which means that every tensor product pair needs to be a candidate for swapping, and as it can be seen from the result, they are all there, on the other side, as an output. By taking into account amplitude values and the equality thereof, the probability of observing a particular swapped state as a result also needs to be equal, which it is. If we take, for example, the amplitude state $ \ket{01} $ with the amplitude of $ \frac{1}{2} $, by squaring the amplitude and thus obtaining the probability of observing that characteristic, we have $ \frac{1}{4} $--and that is exactly what we expect as per our own reasoning, as input and output are linked.

In previous examples, we were performing calculations in a single sequence, but it is also possible to perform them in parallel and then, at some point, "merge" results and continue, for example, in a single sequence. How an algorithm will look depends on the problem and the designer of the algorithm. By constructing a quantum circuit, one can manipulate events and, in turn, the probability of amplitude states, transforming input into output and a problem into a solution. Therefore, with that in mind, we will in the continuation show the often-used algorithm design pattern useful to solve various quantum conundrums, namely the Bernstein-Vazirani design pattern.

\subsection{Bernstein–Vazirani Algorithm Design Pattern}
\label{sec:BVADP}

One might think that quantum computers have an upper hand over classical computers in terms of computability; however, this is not the case. \cite{Vamos2011} Every problem that a quantum machine can solve can also be solved on a classical computer, thus not making a quantum machine superior in that respect; as a consequence, problems that are undecidable in a classical case, which are the hardest problems in existence \cite{Nielsen2010}, are also undecidable for quantum computers. \cite{Vamos2011} What makes quantum computers of interest are superposition, quantum parallelism, and entanglement, as these make quantum machines perform faster. \cite{Nielsen2010,Vamos2011,AHARONOV1999}

"The heart of any quantum algorithm is the way in which it manipulates quantum parallelism so that desired results will be measured with high probability." \cite{Rieffel2000} What brings us to the Bernstein-Vazirani algorithm, which uses superposition, quantum parallelism, and an effect called phase-kickback, so as to achieve its result. \cite{Bernstein1997} These manipulations have no analog in the classical computer world; therefore, a quantum computer is necessary to bring the aforementioned algorithm into reality. \cite{Rieffel2000}

There is a problem of determining the value of each character in a string. \cite{Naseri2022} For example, one might have the following string, $ 1100 $. The question then is: what algorithm could we devise in order to determine in which place the string has a raised bit? As the reader might already guess, we would need to perform a logical conjunction for every bit, as presented in the following equation. \cite{Bernstein1997,Naseri2022}

\begin{equation}
	\begin{matrix}
		& 1 & 1 & 0 & 0 \\
		\& &&&&1\\\hline
		&&&& 0
	\end{matrix}{\mkern 30mu}
	\begin{matrix}
		& 1 & 1 & 0 & 0 \\
		\& &&&1&\\\hline
		&&&0&
	\end{matrix}{\mkern 30mu}
	\begin{matrix}
		& 1 & 1 & 0 & 0 \\
		\& &&1&&\\\hline
		&&1&&
	\end{matrix}{\mkern 30mu}
	\begin{matrix}
		& 1 & 1 & 0 & 0 \\
		\& &1&&&\\\hline
		&1&&&
	\end{matrix}
\end{equation}

\begin{figure}[h]% Bernstein–Vazirani
	\centering
	\includegraphics[width=1\textwidth]{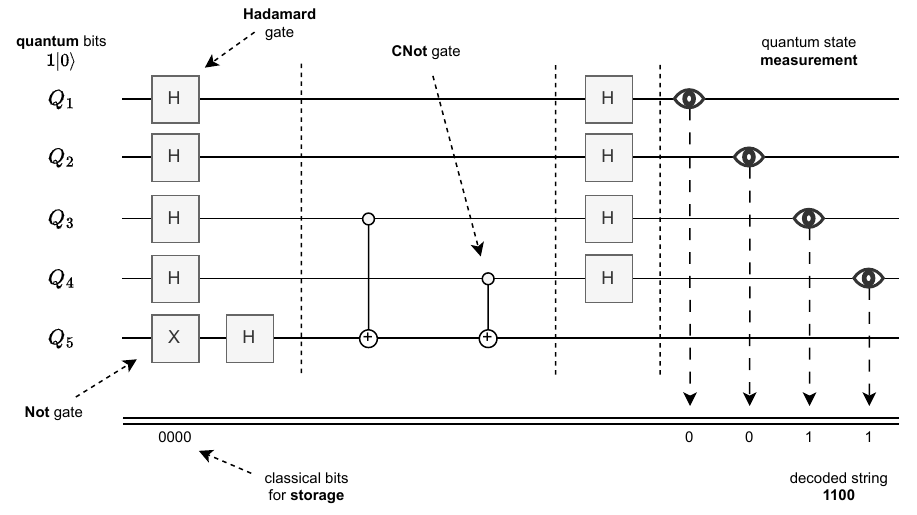}
	\caption{Quantum Circuit of the Bernstein-Vazirani Algorithm. \cite{Bernstein1997,Naseri2022} The horizontal line and horizontal dashed line represent the quantum circuit sequence and operations delimiter, respectively. Gate denoted as X is also known under the name PauliX, as it makes a rotation around the $ X $ axis by $\theta = \pi$ radians. A circle without the plus symbol of the CNot gate represents a control, while a circle with the plus symbol denotes a target. Two parallel lines at the bottom of the figure represent classical storage necessary for saving a result of the quantum algorithm. A quantum algorithm circuit looks like a sheet of music note paper, and there is some resemblance--we are playing a magnificent instrument called nature.}\label{fig:BVA}
\end{figure}

And now, by reading from the back, we have the result, $ 1100 $, the original bit string is decoded. It is clearly seen from the example that for a $ n $-bit string, we would need $ n $ operations to find the source bit string--that is, with the linear increase of the input, the complexity of the algorithm increases linearly. This is not an inefficient algorithm; however, for a bit string of length $ 10^9 $ the number of steps needed to be performed is substantial, and this is where a quantum computer can excel. By employing characteristics that a quantum machine would have, the aforementioned algorithm could be adapted and the entire calculation done in only one step, and thus regardless of the input string, if the quantum machine can match the problem, the calculation would be completed in one step only--this is outstanding, and the procedure that accomplishes the aforementioned is called the Bernstein-Vazirani algorithm; for a visual representation, one can consult Figure~\ref{fig:BVA}.

Before we perform some calculations, let us expound on a fundamental idea behind the Bernstein-Vazirani algorithm. Since our string is four bits long, we also need four qubits for the quantum algorithm as well. The quantum algorithm, however, needs one additional qubit through which the essence of the algorithm will be delivered. All the qubits are at the beginning in the ground state of $ \ket{0} $. These qubits are then placed in a superposition of values, while the last qubit is first placed in a $ \ket{1} $ and then into a superposition, which means that the last qubit has a phase added to its superposition, and this is crucial.

In the next series of operations, there are CNot gates added to every qubit on which we need to decode 1, an excited state, with the last qubit, a qubit with a phase in its superposition, being a target of the CNot. This part of the algorithm is the part where the flash happens, as the phase from the target qubit transfers onto the control qubits, a target has had an influence on the control; this unexpected event is known under the name phase-kickback \cite{OssorioCastillo2023} and is a crucial part of the algorithm. When we, after this step, perform an additional step with the Hadamard gate and return qubits out of superposition, the phase-kickback will have, as a consequence, a qubit in the state $ \ket{1} $ where before it was $ \ket{0} $. By making measurements on qubits, as a last step of the algorithm, we will read the final state and receive the desired result of the decoded string. This read data is then stored on a classical storage. And so, by using a phase-kickback effect, we were able to detect a desirable characteristic and make a transformation by which the end result was obtained. \cite{OssorioCastillo2023,Bernstein1997}

By performing actual calculations, it can be more clearly seen why this has happened and what the algorithm's inner workings are. At the very start of the algorithm, we need to place qubits into superposition, and as we have already shown this in Equation~\ref{eq:hadCalc}, and as it is quite clear what will happen by applying the Not operation from Equation~\ref{eq:not}, these steps will be skipped. Suffice to say, Hadamard gate will produce, $ \ket{\psi_{1,2,3,4}} = \frac{1}{\sqrt{2}}\ket{0}+\frac{1}{\sqrt{2}}\ket{1} $, the X gate will produce, $ \ket{\psi_{5}} = 0\ket{0} + 1\ket{1}$, and the Hadamard applied after the X gate will produce, $ \ket{\psi_{5}} = \frac{1}{\sqrt{2}}\ket{0}-\frac{1}{\sqrt{2}}\ket{1} $--with this, the first series of operations is finished, and now we are onto phase-kickback.

In the second series of steps, the CNot gate is applied to the qubits where we need to decode 1, and so we have $ \ket{\psi_{3,4}} = \frac{1}{\sqrt{2}}\ket{0}+\frac{1}{\sqrt{2}}\ket{1} $ as the control for their respective CNot gate, while we have $ \ket{\psi_{5}} = \frac{1}{\sqrt{2}}\ket{0}-\frac{1}{\sqrt{2}}\ket{1} $ as the target for both instances. By performing the tensor product $ \ket{\psi_{3}} \otimes \ket{\psi_{5}} $ we have the following.

\begin{equation}
	\label{eq:BernVaz}
	\begin{bmatrix}
		1 & 0 & 0 & 0 \\
		0 & 1 & 0 & 0 \\
		0 & 0 & 0 & 1 \\
		0 & 0 & 1 & 0 \\
	\end{bmatrix}
	\begin{bmatrix*}[r]
		\dfrac{1}{2} \\[0.7em]
		-\dfrac{1}{2} \\[0.7em]
		\dfrac{1}{2} \\[0.7em]
		-\dfrac{1}{2}
	\end{bmatrix*}
	= \dfrac{1}{2}\ket{00} - \dfrac{1}{2}\ket{01} - \dfrac{1}{2}\ket{10} + \dfrac{1}{2}\ket{11}
\end{equation}

By applying the CNot gate, probabilities have not changed, and if we were to measure the states now, at this moment, nothing extraordinary would happen. But, if we observe the mixed state more closely, a change of phase has happened, and this is exactly what we wanted; the target has influenced the control, and thus we have, $ \ket{\psi_{3}} = +\ket{0}-\ket{1} $. What brings us to the last step, just before we are ready to measure the result. If we apply the H gate one more time, we will reverse the superposition, yet as we have changed the phase of certain qubits, these will no longer collapse to their original state but to the opposite one. Let us collapse $ \ket{\psi_{3}} $.

\begin{equation}
	\dfrac{1}{\sqrt{2}}
	\begin{bmatrix}
		1 & 1 \\
		1 & -1 
	\end{bmatrix}
	\begin{bmatrix*}[r]
	\dfrac{1}{\sqrt{2}}	\\
	-\dfrac{1}{\sqrt{2}}
	\end{bmatrix*}
	= 0 + 1
\end{equation}

Which makes the finals state, $ \ket{\psi_{3}} = 0\ket{0} + 1\ket{1} $, a $ 100\% $ chance of observing 1--by which the desired result was obtained, the binary string was decoded, and the information now only needs to be recorded, an operation conducted via a classical storage device. The string $ 1100 $ was the one to find, and while the string $ 1100 $ was the one found, the algorithm works well.

Phase-kickback is a mechanism that is often found in quantum algorithms, like, for example, Grover’s \cite{Grover1996,Haverly2021}, and in Deutsch-Josza \cite{Deutsch1992,OssorioCastillo2023}, and it is therefore important to conquer this design pattern. The basic idea behind such algorithms is to develop a quantum "oracle" that will only apply the negative phase to a state one is looking for, which is by no means an easy task to do, and when that is achieved, we can perform, as necessary, amplitude amplification, thus diminishing undesirable amplitudes and increasing the desirable ones, which is the way by which a quantum computer increases the probability of success so as to ensure a sought-after result is measured\footnote{If we have an ion-trap quantum computer implementation, then for a readout, "one illuminates an atom with light of an appropriate frequency so that atoms in the ground state strongly scatter the light, while atoms in the excited state are transparent", and so "by observing whether the illuminated ion glows or not, we can determine with high confidence whether the state of the qubit is $ \ket{0} $ or $ \ket{1} $." \cite{Preskill2021}} with high occurrence probability. \cite{Grover1996,Haverly2021,Deutsch1992,OssorioCastillo2023,Shor1994}

\section{Questions that Puzzle the Mind}
\label{sec:QPM}

Among many intriguing problems in quantum computing that are in need of solving and that are also of interest, there are some that are of special stake for us here. In spite of all the accomplishments, the more reliable and broader reality of quantum computers is still a dream. The main issues standing in the way of quantum computer construction are the following:

\begin{enumerate}
	\item "The possibility in principle to construct a scalable quantum computer." \cite{Savchuk2019}
	\item "Instability (decoherence) because of the influence of external environment." \cite{Savchuk2019}
	\item "A physical implementation of a scalable quantum computer with a sufficient (for practical problems) number of jointly operating qubits\footnote{"IBM releases first-ever 1000-qubit quantum chip--but will now focus on developing smaller chips with a fresh approach to ‘error correction’." \cite{Castelvecchi2023}}." \cite{Savchuk2019}
	\item "The uncertainty of the degree of dependence of errors since a very fast accumulation of errors with increasing the number of qubits will give no way to obtain the sought-for result when executing computations with an acceptable number of repetitions." \cite{Savchuk2019}
	\item "The construction of new mathematical algorithms that will allow to considerably accelerate computations and the search for solutions for a wide class of problems." \cite{Savchuk2019}
\end{enumerate}

Another area of research that is quite significant is finding the position of quantum computation with regard to classical computation in terms of computational cost and complexity classes, as well as exploring the limitations of models of computation. \cite{Qiu2008} As such, there exists a complexity class BQP (bounded-error quantum polynomial time) that consists of decision problems that can be solved by a quantum machine in polynomial time, with the probability of a correct answer being $ \geq \frac{2}{3} $. \cite{Bremner2009,Younes2015} This complexity class is a quantum analogue for the classical BPP (bounded-error probabilistic polynomial time) that "consists of problems for which there exists a polynomial-time Atlantic City\footnote{"Atlantic City randomized algorithm is a probabilistic polynomial-time algorithm that gives a correct answer with the probability $ P_{OPT} $ of at least $ \frac{3}{4} $." \cite{Kudelic2023}} algorithm with a two-sided error." \cite{Kudelic2023} These classes are related in the following way, $ BPP\subseteq BQP $, with both classes belonging to PSPACE and needing a polynomial amount of space. \cite{Kudelic2023} The question of BQP and its relation to NP is a matter that is more difficult. There are indications that perhaps NP is contained in BQP, as there are results for the opposite being true--this then still represents a question that is an issue in science and is considered unresolved. \cite{Younes2015}

Quantum computation is fascinating from yet another perspective, which is the very basis of it, that is because of quantum mechanics. Quantum mechanics is the fundamental theory in physics describing nature at the smallest of scale, at the atomic and subatomic level \cite{Feynman1989}, yet it seems that even quantum mechanics does not give all the answers, and not only for the reason of Gödel's incompleteness theorems \cite{Smullyan2017}. There are certain aspects that escape us, at least for the time being, with entanglement and non-locality representing parts of the picture. \cite{Perlman2017} The issue is, however, broader, as the theory of quantum mechanics fails to address the question of, "how even a single particle, by being in a given quantum state, causes the frequency distribution of measurement values specified by the state." \cite{Perlman2017} And so, the never-ending pursuit in science, for new knowledge and discoveries, continues.

\section{Moving Forward}
\label{sec:Computation}

It is tempting to think that one should use a quantum computer for every problem and for every task; quantum computers, however, are not a key that fits into every lock. There are problems that naturally fit quantum computing and those that do not. The most obvious application of a quantum computer is naturally quantum simulation \cite{Georgescu2014}. By using a quantum computer, one can cope well with the complexity that overwhelms a classical machine. Examples of such modeling include superconductivity \cite{IMADA1990}, chemical processes \cite{Babbush2014}, photosynthesis \cite{Wang2018}, physics processes \cite{Georgescu2014}, cosmology \cite{Liu2021}, etc. Other, more classical examples, so to speak, are cryptography \cite{Pirandola2020}, optimization \cite{Li2020}, search \cite{Yoder2014}, and also machine learning and artificial intelligence \cite{Ciliberto2018,Dunjko2018}.

There are two main types of quantum computer implementation: universal \cite{Lagana2009}, and non-universal \cite{Savchuk2019}. "The main distinction is that universal quantum computing devices are developed with a view to executing arbitrary allowed operations and solving arbitrary problems; while non-universal computing devices are created to solve some limited class of problems, for example, to optimize definite machine learning algorithms." \cite{Savchuk2019}

These quantum machines can be implemented in various ways, with different physical technologies in mind, like trapped ions, superconductors, or photons. \cite{Resch2019} Each individual technology has its ups and downs; in each case, however, quantum computers "are very hard to build"; with the thread that permeates all implementations being quantum noise. \cite{Resch2019} "Quantum mechanical states are extremely fragile and require near-absolute isolation from the environment; such conditions are hard to create and typically require temperatures near absolute zero and shielding from radiation." \cite{Resch2019} Which makes quantum computers expensive to build and difficult to operate. \cite{Resch2019} As the size of a quantum computer increases, so do the challenges, which get mounted one upon the other (in terms of the "number of qubits and the length of time they must be coherent"). \cite{Resch2019}

When computation is being done on a quantum machine, that is, on encoded states, "qubits interact with each other through the gates, and this way errors can propagate through the gates, from one qubit to another." \cite{AHARONOV1999} In such a manner, the error can quickly be spread to all of the qubits. \cite{AHARONOV1999} To solve this problem, computation and error correction can be performed in a distributed way so that "each qubit can effect only a small number of other qubits." \cite{AHARONOV1999} An estimation was made that "more than $ 99\% $ of the computation performed by a quantum computer will be for error correction." \cite{Resch2019,KregerStickles2008} If that is the case, then the calculations that a quantum computer should perform become of secondary nature, thus making the goal of fault-tolerant quantum operations of extremely high importance. \cite{KregerStickles2008} By taking that fact into context, quantum practicality will be a difficult goal to achieve, as a commercial quantum computer would need thousands and millions of qubits--efforts are, however, being made in order to solve the issue. \cite{Hoefler2023,Castelvecchi2023}

Quantum computers have limitations that go beyond their applicability. In spite of having a general scheme for speeding up computation, it is not expected to solve efficiently and in an exact manner NP-hard optimization problems. \cite{Preskill2021,Hey2023} In order to make quantum practicality a reality, significant algorithmic improvements are yet to be achieved, while "due to limitations of input and output bandwidth, quantum computers will be practical for "big compute" problems on small data, not big data problems." \cite{Hoefler2023} Nevertheless, through continuing progress and innovation, it is expected that a quantum computer able to break RSA-4096, with a probability of $ \frac{1}{2} $, will be constructed within the next 10-15 years. \cite{Savchuk2019} With that in mind, it is necessary to already prepare options for replacement so as to ensure post-quantum cryptography viability. \cite{Savchuk2019}

In the meantime, until commercial quantum computers are a reality, it is possible to create variational quantum algorithms that are trying to merge the classical and quantum approaches to problems. \cite{Cerezo2021} In order to deal with the limitations of quantum computers, such as the limit on the number of qubits and the limit on the circuit depth as per noise, a variational quantum algorithm can be used instead. \cite{Cerezo2021} Such an algorithm uses "a classical optimizer to train a parameterized quantum circuit." \cite{Cerezo2021} In spite of the challenges of these algorithms as well, like trainability, accuracy, and efficiency, they are, for the short term at least, perhaps the best option for making the quantum dream a reality in the here and now. \cite{Cerezo2021}

In order to start building quantum algorithms now, the following resources represent possible starting positions. In \cite{Martyn2021} one can read about a quantum singular value transformation (QSVT), which represents a general framework for a number of quantum algorithms, with the possibility of suggesting a unification of quantum algorithms. \cite{Martyn2021} While the following materials represent practical and hands-on foundational experience in quantum computing: \cite{Sutor2019}, \cite{Hidary2021}\footnote{\url{https://github.com/JackHidary/quantumcomputingbook}}, \cite{Matthews2021}, \cite{Intel2024}, \cite{IBM2024}, \cite{Microsoft2024}, \cite{Google2024}.

\section{Few Last Words}
\label{sec:SA}

It was the goal of this research to present to the scientific community an in-depth historical and current survey of quantum computing, with a special emphasis on foundational concepts that are difficult to grasp while also gazing into the future--and almost all of it has been done, from history to terminology, from quantum effects to quantum computation, and from the standard model algorithmics to the related literature. It is therefore left for us to touch upon wrapping issues, consider open questions, and draw conclusions.

Even tough, at times it might seem hopeless that a true, large-scale quantum computer will some day be a reality. Science is advancing, and every year there comes some new experimental success, and this ambitious dream of quantum computation might be possible. \cite{AHARONOV1999,Castelvecchi2023,Waintal2023}

Quantum entanglement is of special interest as it allows for the teleportation of quantum states, and as it is currently known, there is no limit on the distance, which could perhaps enable a large-scale network, a marvel that would be quantum internet. \cite{Resch2019} Considering that quantum encryption can't be broken, even in theory, such a communication network is of great interest and would be of incredible value--it would be the absolute security realized. \cite{Resch2019,Yin2020}

If we have learned anything thus far, it is the fact that realizing a quantum computer, even of any kind, is not an easy task; however, Quantum David just might overpower Classical Goliath. \cite{Preskill2021} By superconducting quantum technology, Google was successful in constructing Sycamore, a programmable quantum machine that has 53 qubits. \cite{Preskill2021} For the reason of errors, "the final measurement yields the correct output only once in 500 runs", yet if one makes repeated calculations "millions of times in just a few minutes", a statistically useful result can be obtained. \cite{Preskill2021} The Sycamore quantum computer is only a single chip, compared to a classical computer that spans tennis courts and uses megawatts of power. \cite{Preskill2021} And Google is not the only one; IBM, for example, paves the way for an error-resilient quantum computer with thousands of qubits. \cite{Castelvecchi2023} Indeed, sufficient progress has still not been achieved in realizing a scalable quantum device, it is nevertheless perceived that, with the developments at hand, "a full-fledged quantum computer will be created in the next 10-15 years." \cite{Savchuk2019}

At the present, quantum mechanics is "considered the most accurate description of the Universe", although the theory might need modifications in the future. \cite{AHARONOV1999,Perlman2017} If and when such a scenario becomes a reality, it is unclear how will that change in the theory of quantum mechanics reflect on quantum computing and quantum information; however, "the novel physical theory that will emerge may give rise to a new computational paradigm, maybe even more powerful than quantum computing." \cite{AHARONOV1999} There is a possibility that large-scale commercial quantum devices won't be feasible, perhaps because of a currently unknown or unsolvable issue--in such a case, a quantum computer can still be useful, e.g. for being "the simulator Feynman first envisaged", or for allowing experimental research in physics, and thus, by manipulating a small number of qubits, physicists will be performing tests and validating predictions of quantum theory. \cite{AHARONOV1999}

Even though it is not expected that quantum computers, via quantum algorithms, will be able to solve NP-complete problems in a manner that is exact and efficient, there is a possibility of finding efficient algorithms for those problems for which we do not know whether they belong to a class of NP-complete problems and do not have known and efficient classical algorithms, like, for example, the problem of "checking whether two graphs are isomorphic, known as Graph Isomorphism\footnote{The current state of the art is the algorithm by László Babai, for which it is claimed to have a quasi-polynomial time. \cite{Babai2017}}." \cite{AHARONOV1999,Preskill2021,Hey2023}

In spite of all of its marvels and all of the scientific contributions, there are many unsolved/partially solved open problems in the realm of quantum computing and quantum mechanics. Here we will list just a small fraction of those, which are likely also the most pressing and fascinating.

\begin{itemize}
	\item[$ \bullet $] Reduction of quantum error rates. \cite{Gyongyosi2019,Krinner2022,Cho2021}
	\item[$ \bullet $] Suppression of quantum decoherence. \cite{Shor1995,Wu2023}
	\item[$ \bullet $] Finding a type of technology best suited for quantum computation and an implementation thereof. \cite{Joshi2024,Yu2023,Zhao2023,Heussen2023,Gschwendtner2023,Psaroudaki2023}
	\item[$ \bullet $] The relationship in regard to NP and BQP. \cite{Bennett1997,Creiner2023}
	\item[$ \bullet $] Scalability of a quantum computer. \cite{FellousAsiani2023,Skoric2023}
	\item[$ \bullet $] Verification of a quantum system. \cite{Gheorghiu2018,Shaffer2021}
	\item[$ \bullet $] Separation of BQP and PH outside of a black-box model. \cite{Raz2022}
	\item[$ \bullet $] Efficient quantum memory. \cite{Rietsche2022,Kimura2023}
	\item[$ \bullet $] Networking protocols and devices for the quantum internet. \cite{Cacciapuoti2020,Azuma2023}
	\item[$ \bullet $] Balance of connectivity between qubits. \cite{Corcoles2020,Yuan2023}
	\item[$ \bullet $] Performance of a quantum gate set. \cite{Corcoles2020,Lacroix2020}
	\item[$ \bullet $] Compilers and software stack performance. \cite{Corcoles2020,Lubinski2023,Cuomo2023}
	\item[$ \bullet $] Materials challenges in quantum computing. \cite{Leon2021,Alfieri2022}
	\item[$ \bullet $] Distributed quantum computing challenges. \cite{Gyongyosi2019,Acampora2023}
	\item[$ \bullet $] Quantum computing programming language challenges. \cite{Voichick2023,Khan2023}
	\item[$ \bullet $] Realizing quantum service-oriented computing. \cite{Moguel2022,Beisel2023}
	\item[$ \bullet $] Efficient, practical, and reliable interface between classical and quantum computers. \cite{Reilly2019}
	\item[$ \bullet $] Quantum machine learning model trainability. \cite{Cerezo2022}
	\item[$ \bullet $] Improvements of quantum algorithms. \cite{Hoefler2023,Lee2023}
	\item[$ \bullet $] Advancing the theory of quantum mechanics and reflecting those findings to quantum computation. \cite{Perlman2017,Soulas2023,Wallace2023}
	\item[$ \bullet $] Solving new moral and social problems raised by quantum computation. \cite{Possati2023,TenHolter2021}
\end{itemize}

Alongside the previous literature list corresponding to a number of quantum computing open problems, one could also consult the following literature as well, \cite{Horodecki2022,Aaronson2021,Shubham2019,Biswas2017,Ho2018,Khrennikov2014,Zinner2022,Huang2017,Ronagh2019,Cumming2022,Pouse2023}, while for the skeptic's view on quantum computation, the following IEEE article is a good read, \cite{Gent2023}. For an article that might be a valuable resource for anyone wanting to continue his quantum journey, so that the beginning of your quantum journey, if that is the case, won't be the beginning of the end, a somewhat older but still contextually relevant article for a non-physicist can be found in the ACM's digital library, \cite{Rieffel2000}; with the quantum algorithm implementations being presented in the following material, in \cite{Abhijith2022} and \cite{Montanaro2016}. More advanced topics on quantum computing can easily be found in the article's references; an advanced expert will no doubt manage its course.

The question of the importance of quantum physics and its future practical prospects is debated, some say that we are in a second quantum revolution where "you’re engineering the quantum mechanics itself to do something", while others are still doubting that there will ever be anything serious enough for large-scale application. \cite{NIST2022,AHARONOV1999,Gent2023} Whatever it may be, a brick wall has not been hit yet, and the race is on: "from nations to corporations, everyone is getting into the game" \cite{NIST2022}, and just as information can't be created nor deleted due to the conservation of quantum information \cite{Zyga2011}, similarly, the will to succeed in quantum computing still holds strong. Many new discoveries await, some as inventors, some as authors, some as readers, and some as users. The best is indeed yet to come, an optimist would claim, and why should we not be optimistic, one could ask. Thus, on a more personal note to the reader, if I may, I wish you a most prosperous race.

%\backmatter

%\bmhead{Supplementary information}
%
%If your article has accompanying supplementary file/s please state so here. 
%
%Authors reporting data from electrophoretic gels and blots should supply the full unprocessed scans for key as part of their Supplementary information. This may be requested by the editorial team/s if it is missing.
%
%Please refer to Journal-level guidance for any specific requirements.
%
\section*{Acknowledgments}

For the research, the following was also used: Latex\footnote{https://www.latex-project.org/}, TexLive\footnote{https://tug.org/texlive/}, Draw.io\footnote{https://www.drawio.com/}, Texstudio\footnote{https://www.texstudio.org/}, LibreOffice\footnote{https://www.libreoffice.org/}, JabRef\footnote{https://www.jabref.org/}, and Linux Mint\footnote{https://linuxmint.com/}.

\section*{Declarations}

The author declares no conflict of interest.

\clearpage
\bibliographystyle{splncs04}
\bibliography{mybibliography}
%Author, F.: Article title. Journal \textbf{2}(5), 99--110 (2016)
%Author, F., Author, S.: Title of a proceedings paper. In: Editor,
%F., Editor, S. (eds.) CONFERENCE 2016, LNCS, vol. 9999, pp. 1--13.
%Springer, Heidelberg (2016). \doi{10.10007/1234567890}
%Author, F., Author, S., Author, T.: Book title. 2nd edn. Publisher,
%Author, A.-B.: Contribution title. In: 9th International Proceedings
%on Proceedings, pp. 1--2. Publisher, Location (2010)
%LNCS Homepage, \url{http://www.springer.com/lncs}. Last accessed 4
%\end{thebibliography}
\end{document}